\documentclass[sigconf,nonacm]{acmart}

\copyrightyear{2026}

\usepackage{booktabs}
\usepackage{multirow}
\usepackage{graphicx}
\usepackage{xcolor}
\usepackage{subcaption}
\usepackage{balance}
\usepackage{pifont}
\usepackage{siunitx}
\usepackage{tikz}
\usepackage{enumitem}
\usetikzlibrary{arrows.meta, positioning, shapes.geometric, fit, decorations.pathreplacing, calc, backgrounds}

\newcommand{\system}{Sieve}
\newcommand{\numqueries}{133}
\newcommand{\numlogtypes}{5}

\begin{document}

\title{Parser-Free Querying of Security Logs}

\author{Evan Luo}
\affiliation{%
  \institution{University of California, Berkeley}
  \city{Berkeley}
  \state{California}
  \country{USA}
}
\email{evanluo@berkeley.edu}

\author{Julien Piet}
\affiliation{%
  \institution{University of California, Berkeley}
  \city{Berkeley}
  \state{California}
  \country{USA}
}
\email{julien.piet@berkeley.edu}

\author{David Wagner}
\affiliation{%
  \institution{University of California, Berkeley}
  \city{Berkeley}
  \state{California}
  \country{USA}
}
\email{daw@cs.berkeley.edu}

\begin{abstract}
Security analysts routinely query system logs to detect threats and investigate incidents, but each log source uses its own semi-structured format: logs are cheap to produce, but expensive to use. The standard approach, building per-source parsers to normalize logs into structured schemas, is powerful but requires continuous engineering effort for each new format. Querying raw logs directly with tools like \texttt{grep} avoids this cost, but requires analysts to know each source's message variants and cannot express the multi-line temporal queries that security investigations demand. We present \system{}, a system that generates executable query code from natural-language security questions by grounding a large language model with lightweight, automatically extracted log-format context, requiring only one LLM call per query followed by deterministic execution. Evaluating \numqueries{} security queries across \numlogtypes{} log types, we find that \system{} achieves over a 3$\times$ reduction in error rate on complex temporal and cross-event queries compared to manual analyst scripting, with the largest gains on the multi-line correlation tasks most critical to active investigations. Our results and benchmark provide evidence that LLM-generated code can bridge the gap between the expressiveness of structured log querying and the immediacy of working directly with raw files.
\end{abstract}

\maketitle

\section{Introduction}
\label{sec:intro}

Security operations centers process large volumes of log data from firewalls, authentication services, network infrastructure, and endpoint agents to detect threats and reconstruct incidents. A single investigation may span audit trails, SSH authentication records, DHCP lease transactions, and configuration management state, each recorded in a different semi-structured format. The standard architecture for handling this diversity converts raw logs into structured events before analysis: SIEM platforms ingest each source through a per-source parser, map the resulting fields into normalized schemas such as the Open Cybersecurity Schema Framework (OCSF)~\cite{ocsf}, and expose the result through a query language. When this pipeline works, analysts can ask security-relevant questions (failed logins by source IP, privilege escalations by host, anomalous lease patterns by subnet) without knowing the conventions of every log source.

The difficulty is that parsers perform a large amount of source-specific work. Logs are neither unstructured strings nor clean tables: they mix structured fragments with natural-language text, and each logging source (an application version, a uniquely configured system daemon, a vendor-specific integration) uses its own dialect, producing many distinct templated events. Identifiers vary from one line to another: an SSH failure might appear as ``Failed password for root from \texttt{X}'', ``authentication failure; user=root rhost=\texttt{X}'', or a PAM message with the relevant user and host split across key-value pairs. Building a parser that covers these cases requires encoding both syntax and semantics for each source and maintaining that encoding as formats evolve. In production, this becomes a major continuous engineering task rather than a one-time setup cost. Google SecOps alone maintains roughly 1{,}000 active parsers and pushes more than 200 parser updates each month, requiring more than 4{,}000 engineer-hours per year~\cite{piet2025matryoshka}. Recent log-parsing systems such as LILAC~\cite{jiang2024lilac} and Matryoshka~\cite{piet2025matryoshka} reduce this human effort, but they remain expensive pipelines that either require substantial validation before producing an actionable parser, or must run offline on each new source before that source can be queried. This makes it difficult to query arbitrary logs during an active investigation.

The usual alternative is to skip parsing and search the raw file directly with tools such as \texttt{grep}, \texttt{awk}, and short Python scripts. This approach is attractive because it works immediately on any file, but it shifts the burden back to the analyst. To find failed SSH logins, the analyst must know every phrase that denotes failure in that particular source; to count DHCP requests, they must know that requests can be logged as incoming client messages, server-side lease handling, or relay activity. Even when the right phrases are known, many of the queries most critical to network security cannot be expressed as substring filters: ``find users who logged in from two different IP addresses within 60 seconds'' (a lateral-movement indicator), ``find DHCP transactions where request-to-ack latency exceeds 5 seconds'' (a rogue-server signal), and ``count failures in any rolling 5-minute window'' (brute-force detection) all require state, timestamps, grouping keys, and aggregation.

These two paths fail for the same reason. Whether an analyst writes a durable parser or an ad hoc \texttt{grep} command, the query depends on knowledge of the source log's syntax and semantics: which message formats exist, which tokens vary, which values identify the same entity across lines, and which phrases are equivalent for the query at hand. Structured pipelines encode this knowledge ahead of time in parsers and schemas; raw-text workflows require the analyst to reconstruct it manually during the investigation. Both approaches require source-specific knowledge upfront: parsers are powerful but expensive to write and maintain, and grep-style search shifts that cost to query time, where it remains brittle and limited in expressiveness.

We aim to combine both approaches: immediate support for arbitrary log formats, the expressiveness of structured queries, and no requirement that the analyst know the details of all log formats. Security logs differ from arbitrary unstructured text in one useful way: they contain repeated formats and recurring identifiers, making it possible to summarize the space of relevant variations without showing the model the entire file. In practice, a log file usually contains a finite and manageable set of message templates. Across all possible log sources the number of templates is unbounded; within a single source it remains small enough to summarize, with under 500 templates in our datasets and similarly bounded counts for sources already supported by SIEM parser libraries. This follows from how logs are generated: source programs contain a finite number of format strings, and runtime records instantiate those strings with different values. Writing a query requires knowing these formats, a task tedious for a human analyst to perform manually but one that can be automated and exposed to a language model.

In particular, we develop AI-powered methods to search log files efficiently. Analysts can specify their query in natural language. A language model then intelligently writes code to scan the logs and answer the analyst's query, based on the formats of relevant log messages. Instead of writing per-source parsers, we propose to (1) extract unique templates from log streams using algorithms that run in seconds to about a minute on logs up to several hundred thousand lines, (2) provide examples of those templates to an LLM along with the analyst's query, (3) use the LLM to identify the formats and templates relevant to the analyst's query, and (4) have the LLM write code that scans the file, finds all relevant log entries, extracts the required values, and computes the answer. The result is a just-in-time query plan over raw text: the analyst writes a high-level question, the LLM writes Python or shell code, and the code executes deterministically on the complete log.

Consider a concrete security query: ``find source IPs with 10 or more SSH authentication failures within any 5-minute window,'' a standard brute-force detection rule. A parser-backed SIEM can answer this only if the event is representable in the target taxonomy and if the parser maps all relevant SSH failure variants into the expected fields for failure events, source addresses, users, and timestamps. Neither condition is guaranteed: an ``SSH authentication failure'' may be expressed by several message families, and the normalized schema may not preserve the distinction or correlation key the analyst needs. If either condition fails, the analyst must extend the taxonomy mapping, extend the parser, or bypass the structured pipeline entirely. A manual approach must enumerate the failure phrases, extract IP addresses from several line shapes, parse syslog timestamps, group by address, and maintain a rolling window without accidentally counting unrelated authentication messages, a task that requires understanding a multitude of log templates and writing a non-trivial ad hoc script. In our approach, the prompt gives the LLM the query together with examples or templates of the SSH message formats. The generated script performs the tedious source-specific work (matching failure variants, extracting timestamps and addresses, and sliding the window) while the analyst can focus on what they're looking for without worrying about tedious implementation details.

Our system, \system{}, represents a deliberately constrained use of LLMs. We do not ask the model to read every line, classify every event, or serve as the database; LLMs are too slow and too expensive for per-line inference over production logs. Instead, \system{} uses the model as a compiler: we generate code from the analyst's prompt, and the code performs the scan, join, and aggregation at normal program speed. Once generated, the execution is deterministic, inspectable, cacheable, and scales with log size like any other script. LLM usage remains affordable because we do not process the entire log with an LLM.

This paper makes four contributions. First, we present \system{}, an open-source pipeline for LLM-assisted querying of raw security logs. Given a natural-language query, \system{} assembles a prompt from the query, optional log templates, and a reservoir-sampled subset of log lines; sends the prompt to an LLM (we run our experiments primarily on Gemini~3.1 Pro Preview~\cite{gemini2025} and OpenAI GPT-5.4~\cite{gpt54systemcard}, with additional Gemini~3 Flash Preview~\cite{gemini3flash2026} results in the appendix); executes the returned Python or Bash code in a sandboxed subprocess with no network or filesystem access; and automatically retries with error feedback up to four times on execution failure, output-format violation, or safety-check rejection.
Second, we construct a benchmark that researchers can use for evaluating solutions for this problem.
The benchmark contains \numqueries{} queries, spanning \numlogtypes{} security-relevant log types (Linux Audit, SSH, cron, Puppet, and DHCP), including both single-line filter queries and complex multi-line queries that require correlating events across lines, parsing timestamps, and computing aggregations.  We also provide ground truth correct answers for each query.
Third, we explore the design space for just-in-time log querying (how to provide log-format information to the model, which execution language to generate, which model is used, and which failure modes arise). We evaluate Matryoshka-generated templates~\cite{piet2025matryoshka}, Drain3 templates~\cite{he2017drain}, a frequency-based templater we introduce, random raw samples, and no context, across two model vendors. Fourth, we compare against a human baseline of \numqueries{} naive \texttt{grep}-and-Python scripts representing the kind of code a SOC analyst would write in under ten minutes during an active investigation, and against an automated structured-parsing pipeline~\cite{piet2025matryoshka}, evaluated on the same SecurityLogs corpus.

Our central finding is that LLM-generated code substantially outperforms manual analyst scripting on complex security queries. \system{} matches the human baseline on simple filter queries, and on the complex workloads (brute-force burst detection, session anomaly identification, cross-event correlation, and temporal-window analysis) it reduces the average error rate from 0.34 to 0.10, a roughly 3.6$\times$ reduction. The gains are largest on the queries where manual scripting is weakest: those that require parsing timestamps across line formats, joining events by shared identifiers, and maintaining rolling-window state. \system{}'s results are obtained with a templater that runs in seconds to under a minute with no LLM calls, so the per-query cost of the system is bounded by a single LLM call followed by deterministic execution.

A second finding is that the choice of template source matters less than expected. A heuristic frequency-based templater that runs in under ten seconds on each of our datasets, with no LLM calls or human review, produces statistically indistinguishable results from Matryoshka templates (the highest-effort, highest-quality template source we evaluate, generated by an offline LLM pipeline) across $N{=}5$ repeated runs. What matters is coverage of the source log's message variants, not the sophistication of the template representation. This indicates that we do not need sophisticated mechanisms to find examples of all message variants relevant to the analyst's query; simple mechanisms suffice.

The remainder of the paper develops this argument. \S\ref{sec:background} describes the structure of security logs and the query workloads we target. \S\ref{sec:related} positions \system{} relative to log parsing, LLM-based log analysis, and natural-language-to-code systems. \S\ref{sec:system} presents the pipeline and context strategies. \S\ref{sec:eval} describes the datasets, ground truth, and baselines. \S\ref{sec:results} reports the empirical results, and \S\ref{sec:discussion} discusses deployment implications and limitations. We release all code, ground-truth parsers, and human baseline scripts to support reproducible research.

\section{Background}
\label{sec:background}

This section provides background on \system{}, covering security log structure (\S\ref{sec:anatomy}), how security operations teams use them (\S\ref{sec:soc}), prior AI-assisted approaches and motivation for our design choices (\S\ref{sec:llm-codegen}), and the target query workloads (\S\ref{sec:queryclasses}).

\subsection{Security logs}
\label{sec:anatomy}

Security logs are records emitted by programs to capture events of interest. Each record carries fixed tokens that identify the event and variable slots that carry its parameters. The dominant format in practice is semi-structured plain text produced by template interpolation: an sshd ``Accepted'' event, for example, is generated by code equivalent to \texttt{printf("Accepted \%s for \%s from \%s port \%d", method, user, ip, port)}, yielding lines such as \texttt{"Accepted publickey for root from 10.35.161.71 port 59271"}. The template is not made explicit in the log, so the analyst must recover it from examples. Fully structured variants also exist, most notably JSON-formatted logs emitted by cloud services, container runtimes, and modern application frameworks. Field names are explicit there, but schemas still vary across sources and unstructured message strings are routinely embedded inside JSON fields, so many of the same querying challenges apply. \system{} treats both forms uniformly because its context extraction and code generation operate on raw log content rather than on a presumed delimiter convention. Throughout this paper we focus on the semi-structured plain-text case. Log variables encode security-critical semantics, such as the user and source IP in the SSH example, and many queries reduce to extracting and aggregating them.

Four properties of security logs complicate automated querying. First, \emph{format variation across sources and versions}: each program defines its own templates, and templates evolve across software versions; extraction logic that works on one source or release silently fails on another. Second, \emph{mixed structure}: within a line, structured delimiters (\texttt{=}, \texttt{:}, braces, quoted strings) interleave with free text, so field extraction by position or simple delimiter splitting misaligns across templates and returns wrong values rather than errors. Third, \emph{template multiplicity}: a single source can contain hundreds of distinct templates (in our datasets, Audit 496, Puppet 254, DHCP 178, sshd 98, and cron 7~\cite{piet2025matryoshka}). A query that targets a semantic event must enumerate every variant; missed variants drop silently from the answer. Fourth, \emph{variable ambiguity}: the same concept (source IP, process ID, user name) appears under different field names and positions across templates and sources~\cite{piet2025matryoshka}, so filtering or joining on a single field name misses every occurrence expressed under a different label.

\subsection{Security operations}
\label{sec:soc}

Security operations teams use logs in two recurring workflows: \emph{interactive investigations} that reconstruct chains of events during incident response, and \emph{real-time detection} that evaluates rules against live log streams to flag suspicious activity. Both workflows depend on the same primitive: a query mechanism that extracts semantically meaningful events (e.g., who did what to whom, when, and from where) from the underlying log stream. The cost of that primitive is operationally significant: the IBM/Ponemon 2024 Cost of a Data Breach study reports that breaches take an average of 194 days to identify, and that organizations with extensive automation in their detection pipelines identify and contain incidents 98 days faster, saving an average of \$2.2M per breach~\cite{ibmcodb2024}. Faster log querying directly shortens this timeline. Our ground-truth query sets (\S\ref{sec:datasets}) are drawn from the canonical SOC task categories codified in the NICCS NICE cybersecurity workforce framework~\cite{niccs}, including resource access analysis, network reconnaissance, spam monitoring, and user-management auditing.

Industry data illustrates the scale of this querying problem. The 2025 SANS SOC Survey finds that 85\% of SOCs trigger incident response primarily from endpoint alerts rather than proactive log-based detection, and that 42\% of organizations ingest all data into a SIEM with no plan to retrieve or analyze it~\cite{sans2025soc}. Much of the security telemetry organizations collect is therefore never effectively queried. This is not because the data lacks value, but because the tooling to extract that value is too expensive or too rigid.

This rigidity stems from how the industry organizes log access. The dominant architecture is \emph{normalize first, query later}: SIEM platforms ingest raw logs through per-source parsers that extract fixed fields into a structured schema, and analysts then query the normalized representation through a platform-specific query language. Adopting this architecture means committing organizational effort to writing and maintaining parsers for each new source and to learning the platform's taxonomy, both of which are substantial barriers; many operators consequently leave their logs unparsed and unqueried. The maintenance cost is visible at scale: Google SecOps alone maintains roughly 1{,}000 active parsers~\cite{googlesecopsparsers} and ships more than 200 parser updates each month~\cite{piet2025matryoshka}, reflecting the continuous churn of log formats in production environments; Splunk's supported add-ons register a similar volume of log sources~\cite{splunkaddons}, and Elastic Security follows a comparable model~\cite{elasticsecurity}. Once a parser is in place, queries execute efficiently; building and maintaining the parser is the bottleneck.

Normalization also imposes an expressivity ceiling. Target taxonomies such as UDM (roughly 20{,}000 attributes~\cite{udm}) and OCSF (over 50{,}000~\cite{ocsf}) are broad, but the binding problem is not their size: it is the misalignment between the fields that the taxonomy exposes and the fields that the source log actually contains, so even fully populated mappings frequently lose information that the original record carried~\cite{piet2025matryoshka}. For queries that reference fields the taxonomy does not represent faithfully, or that require joins the SIEM vendor did not anticipate, the analyst is forced to either (i) extend the parser and wait for redeployment, (ii) fall back to substring search against raw logs, or (iii) write a one-off script to extract the needed data. Each path carries well-known drawbacks: parser extension is slow, substring search is brittle, and one-off scripts must be rewritten for each new log source. \system{} offers an alternative to the entire normalize-first architecture: rather than mapping each source into a static taxonomy, it generates query-specific extraction code on demand from the raw log, removing both the parser-maintenance burden and the taxonomy ceiling while preserving the accuracy of structured querying. This alternative is enabled by recent advances in AI-assisted code generation, which we discuss next.

\subsection{AI for Security Operations}
\label{sec:llm-codegen}

The bottleneck identified above (hand-written parsers and queries written against a static taxonomy) has motivated several lines of AI-assisted work, each of which removes part of the manual cost while leaving the rest in place. We review them in turn before introducing the design used by \system{}.

\paragraph{LLM-based log analysis.}
A first line of work invokes a large language model directly on log data, calling the model on individual lines or small batches to identify events, extract fields, or evaluate detection rules~\cite{beck2025logparser, zhong2024logparserllm, ma2024librelog}. This removes the need for a hand-written parser, but the per-line (or near-per-line) invocation pattern does not scale to the volumes that security telemetry generates and inherits the non-determinism of model output, which is incompatible with detection workflows that demand reproducible results.

\paragraph{LLM-assisted parser generation.}
A second line of work uses LLMs offline to generate the parsers themselves. Matryoshka~\cite{piet2025matryoshka} prompts an LLM with sample lines and obtains templates, field schemas, and taxonomy mappings, which are compiled into deterministic regex-based parsers that run at ingest time with no further LLM calls. This dramatically reduces the manual cost of supporting new log sources, but it preserves the normalize-first architecture and its drawbacks: the resulting parser is still tied to a fixed schema, must be regenerated when the source format evolves, and queries against it must still be written by hand against whatever fields the schema exposes.

\paragraph{LLMs as query compilers.}
The recent advance that makes \system{} viable is that LLMs have become competent code generators from natural-language specifications~\cite{chen2021codex, gemini2025}. This enables a third option that is the foundation of \system{}: rather than asking the model to read every log line or to predefine a parser, we use it as a one-shot query compiler that emits an executable extraction script for each natural-language question. Three properties of this design address the limitations of the approaches above:
\begin{itemize}[leftmargin=*]
    \item \emph{Schema-free operation.} The script reads the raw log directly and recovers only the fields the question requires, removing the dependence on a pre-materialized schema and on its alignment with the source format. Schema-free operation removes the residual cost of parser-generation approaches, in which queries remain bound to a fixed schema.
    \item \emph{Per-query flexibility.} Because code is generated fresh for each query, the expressivity ceiling is the programming language itself rather than a vendor taxonomy. Complex temporal joins, nested aggregations, and derived metrics are as accessible as simple filters, whereas the normalize-first approach forces analysts to fall back on substring search or one-off scripts to express.
    \item \emph{Auditability.} Unlike end-to-end LLM retrieval systems, the generated code is inspectable, deterministic once generated, and can be cached for repeated queries. Determinism is what disqualifies per-line LLM invocation from production detection workflows; the inspectable code path restores it.
\end{itemize}

The cost of this design is one LLM invocation per query (seconds to tens of seconds), in exchange for deterministic execution over the entire log without any per-line model calls. This places \system{} alongside Matryoshka as a complementary point in the LLM-assisted design space: Matryoshka pays its LLM cost once, offline, in exchange for a static parser that suits standing detection rules; \system{} pays it once per query, online, in exchange for a query-specific extractor whose expressivity is not bounded by a parser's schema. A production deployment can plausibly use both. Matryoshka-style parsers can be used for high-volume standing rules and \system{}-style per-query code generation for the long tail of ad hoc queries.

\subsection{Classes of log queries}
\label{sec:queryclasses}

With the structure of security logs and the operational setting in place, we can describe the workloads that \system{} targets. To make the discussion precise, we mirror the standard relational decomposition used in SQL and define four query classes of increasing expressive power, each obtained from the previous one by adding a single operator. Treating the log as a relation $L$ in which each tuple is one record, we consider:

\begin{itemize}[leftmargin=*]
    \item \emph{Class 1 -- selection only (\texttt{WHERE}).} Return the records of $L$ that satisfy a predicate. Example: ``find sshd authentication failures.'' Simple keyword predicates of this form can be handled by line-level filters such as \texttt{grep}, but even Class 1 queries become difficult when the predicate references derived values. For instance, ``find authentication failures between 11{:}00 and 15{:}00'' requires parsing timestamps that may appear in several formats within a single source.
    \item \emph{Class 2 -- projection over selection (\texttt{SELECT}~+~\texttt{WHERE}).} Return one or more named fields from the records that satisfy a predicate. Example: ``list the source IPs that produced sshd authentication failures.'' Class 2 adds a variable extractor on top of the predicate.
    \item \emph{Class 3 -- grouped aggregation (\texttt{SELECT}~+~\texttt{WHERE}~+~\texttt{GROUP BY}).} Return per-group summaries of projected fields restricted to records that satisfy a predicate. Example: ``count failed logins per source IP.'' Class 3 adds stateful accumulation over the projected fields.
    \item \emph{Class 4 -- correlations across records (\texttt{JOIN}).} Return tuples obtained by joining records of $L$ with each other under a key and an additional constraint, typically a temporal one. Examples: ``for each source IP, count failures within any 5-minute window,'' or ``find users who logged in from two different source IPs within 60 seconds of each other.'' Class 4 introduces self-joins on shared identifiers and temporal predicates over timestamps.
\end{itemize}

In our benchmark (\S\ref{sec:datasets}), Class 1 and Class 2 queries together form the \emph{simple} tier, while Class 3 and Class 4 form the \emph{complex} tier; we report results separately by class where the distinction is informative. Throughout the paper, we say that a query is \emph{correct} when the set of records it returns matches the semantically intended answer. Our precise scoring procedure can be found in \S\ref{sec:eval}.


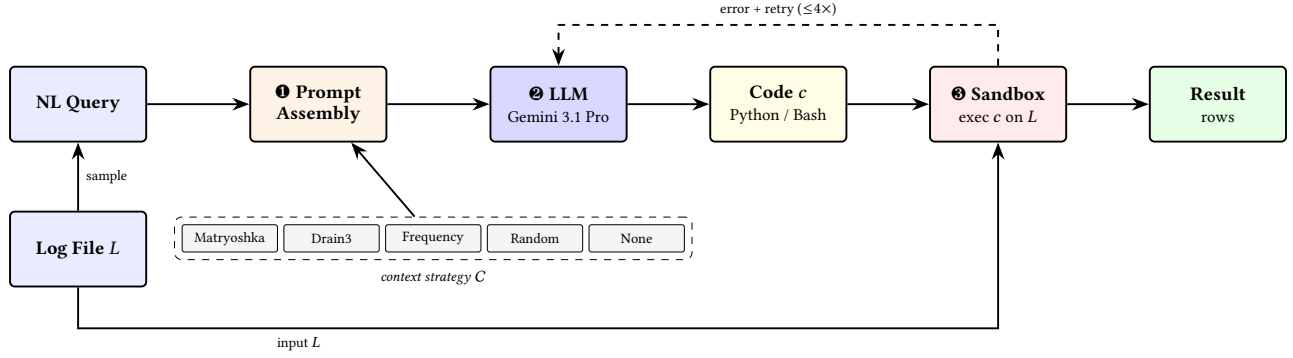
\begin{figure*}[t]
\centering
\resizebox{0.95\textwidth}{!}{%
\begin{tikzpicture}[
    >={Stealth[length=2.5mm]},
    every node/.style={font=\small},
    box/.style={draw, thick, rounded corners=2pt, minimum height=1.1cm,
                minimum width=2.0cm, align=center, inner sep=4pt},
    tag/.style={draw, thin, rounded corners=1pt, minimum height=0.4cm,
                minimum width=1.4cm, align=center, inner sep=2pt,
                font=\scriptsize, fill=gray!8},
    arr/.style={->, thick},
    darr/.style={->, thick, dashed},
]

\node[box, fill=blue!8] (query) {\textbf{NL Query}};

\node[box, fill=orange!10, right=1.5cm of query] (prompt)
    {\ding{182}\;\textbf{Prompt}\\[-1pt]\textbf{Assembly}};

\node[box, fill=blue!15, right=1.5cm of prompt] (llm)
    {\ding{183}\;\textbf{LLM}\\[-1pt]\footnotesize Gemini 3.1 Pro};

\node[box, fill=yellow!12, right=1.2cm of llm] (code)
    {\textbf{Code}~$c$\\[-1pt]\footnotesize Python / Bash};

\node[box, fill=red!8, right=1.2cm of code] (sandbox)
    {\ding{184}\;\textbf{Sandbox}\\[-1pt]\footnotesize exec $c$ on $L$};

\node[box, fill=green!10, right=1.2cm of sandbox] (output)
    {\textbf{Result}\\[-1pt]\footnotesize rows};


\node[box, fill=blue!8, below=1.0cm of query] (logfile)
    {\textbf{Log File}~$L$};

\node[tag, below=1.2cm of prompt, xshift=-1.3cm] (t1) {Matryoshka};
\node[tag, right=0.08cm of t1] (t2) {Drain3};
\node[tag, right=0.08cm of t2] (t3) {Frequency};
\node[tag, right=0.08cm of t3] (t4) {Random};
\node[tag, right=0.08cm of t4] (t5) {None};
\node[draw, dashed, rounded corners=3pt, inner sep=3pt,
      fit=(t1)(t2)(t3)(t4)(t5)] (strat) {};
\node[font=\scriptsize\itshape, below=1pt of strat]
    {context strategy $\mathcal{C}$};


\draw[arr] (query) -- (prompt);

\draw[arr] (logfile) --
    node[midway, right, font=\scriptsize] {sample}
    (prompt.south -| logfile.north);

\draw[arr] (strat) -- (prompt);

\draw[arr] (logfile.south) -- ++(0,-0.6)
    -| node[pos=0.12, below, font=\scriptsize] {input $L$}
    (sandbox.south);

\draw[arr] (prompt) -- (llm);
\draw[arr] (llm) -- (code);
\draw[arr] (code) -- (sandbox);
\draw[arr] (sandbox) -- (output);

\draw[darr] (sandbox.north) -- ++(0,0.6)
    -| node[pos=0.25, above, font=\scriptsize]
       {error + retry (${\leq}4\times$)}
    (llm.north);

\end{tikzpicture}%
}
\caption{\system{} pipeline.
A natural-language query and a log file $L$ are combined with one of five context strategies
$\mathcal{C}$ into a prompt~(\ding{182}).
An LLM emits filter code~$c$ in Python or Bash~(\ding{183}).
The code processes the log $L$ in a sandbox~(\ding{184}).}
\label{fig:system}
\end{figure*}

\section{Related Work}
\label{sec:related}

\paragraph{Log parsing and template mining.}
Log parsing, the task of recovering the printf-style template from a collection of log lines, has been studied for two decades. Static approaches rely on heuristics to tokenize log lines and identify variable slots~\cite{makanju2009iploml, dai2020logram, tang2011logsig, vaarandi2015logcluster, du2016spell}. Two prominent approaches, Drain~\cite{he2017drain} and its successor Brain~\cite{yu2023brain}, build a fixed-depth prefix tree that assigns each line to a template in a single pass, achieving strong accuracy at streaming speed. LogHub~2.0~\cite{jiang2024evaluation, zhu2023loghub} provides a standardized evaluation framework across 14 log types.

More recent approaches use language models to generate templates~\cite{xu2024divlog, huo2024semparser, zhong2024logparserllm, ma2024librelog}. For example, LILAC~\cite{jiang2024lilac} uses an LLM with an adaptive parsing cache to reduce API calls, outperforming all static approaches on LogHub's template generation metrics.  In the security domain specifically, Vaarandi and Bahsi~\cite{vaarandi2025llmtemplate} study LLM-based template detection on security event logs. All of this work targets \emph{template extraction} as the end goal. Templates themselves do not enable log querying: their variables lack semantic meaning and are often too coarse for precise queries~\cite{piet2025matryoshka}. \system{} uses templates only as \emph{prompt context} for a downstream code-generation task; we include Drain3 and a frequency-based templater primarily to study how template quality affects end-to-end query accuracy, rather than to advance template mining itself.

\paragraph{LLM-assisted security log analysis.}
A small but growing body of work applies LLMs directly to security log workloads. SynRAG~\cite{saju2025synrag} uses retrieval-augmented generation~\cite{lewis2020rag} to translate natural-language questions into SIEM-native query languages such as SPL or KQL, operating over data that has already been parsed and normalized by the SIEM. Karlsen et~al.~\cite{karlsen2024benchmarking} benchmark LLMs for security log interpretation and anomaly detection, and RedChronos~\cite{li2025redchronos} builds an LLM-agent pipeline for insider-threat detection over authentication logs. A separate line of work consumes raw unstructured logs to answer natural-language questions via retrieve-then-read pipelines: LogQA~\cite{huang2023logqa} retrieves relevant lines and feeds them to an LLM reader that emits a textual answer, and HG-InsightLog~\cite{bajpai2025hginsightlog} builds a temporal hypergraph over log attribute--value pairs and uses a personalized PageRank traversal to assemble the context window before the LLM produces a natural-language response. The distinguishing property of \system{} relative to all of the systems above is its operating regime: each of them either assumes a pre-parsed, normalized input, targets a specific downstream task (SIEM query translation, anomaly classification, insider-threat detection), or produces a free-form natural-language answer without an executable, inspectable, deterministic artifact. \system{} instead generates per-query extraction code directly from the raw, semi-structured log; the answer is the set of records the code returns, not the prose the model emits, which preserves auditability and reproducibility once the code has been generated. We do not run a head-to-head comparison against SynRAG-style SIEM-query translators because the comparison is not well-defined for our setting: their output (SPL/KQL) requires the very parser-and-schema layer whose elimination is the goal of our system, so reproducing their inputs would necessitate first building exactly the artifact we aim to avoid. The closest comparable prior work that consumes raw security logs is Matryoshka~\cite{piet2025matryoshka}, which uses LLMs offline to generate deterministic parsers; Matryoshka focuses on log ingestion rather than on querying the resulting data, while \system{} is focused on the queries themselves. We compare against \texttt{grep}/short-script baselines (the analyst's default reach for raw logs) and use the same SecurityLogs corpus as Matryoshka, enabling direct methodological comparison where the scopes overlap. The industrial counterparts to these systems, such as Splunk~\cite{splunk, splunkaddons}, Google Security Operations~\cite{googlesecops, googlesecopsparsers}, Microsoft Sentinel, and Elastic Security~\cite{elasticsecurity}, increasingly ship LLM-based natural-language interfaces, but every shipped interface we are aware of translates the user's question into the platform's native structured query language: Splunk's AI Assistant for SPL produces SPL~\cite{splunkaiassistant}, Gemini in Security Operations produces UDM Search and YARA-L~\cite{geminisecops}, Microsoft Security Copilot's Sentinel plugin emits KQL~\cite{securitycopilotnl2kql}, and Elastic's AI Assistant for Security generates ES|QL~\cite{elasticaiassistant}. The LLM in each case sits above the platform's structured query layer rather than over the raw log, so the result still depends on whatever indexing, parsing, or schema mapping the platform applies before query execution; we are not aware of a method that exposes an LLM-driven query path that runs directly against unparsed logs.

\paragraph{LLM code generation and data querying.}
A broader line of work generates code or queries from natural language: Codex~\cite{chen2021codex} for general-purpose code, text-to-SQL systems such as Spider~\cite{yu2018spider} and DIN-SQL~\cite{pourreza2023dinsql} for relational queries against known schemas, and UQE~\cite{dai2024uqe} for query plans over unstructured document collections. Our setting differs in that no externally provided schema is available (the model must infer structure from examples in the prompt), and the metric of interest is retrieval correctness against a ground-truth answer set rather than the functional correctness of the generated code in isolation.

A separate line of work addresses the reliability of LLM-generated code~\cite{chen2024selfdebug, olausson2024selfrepair, zhang2025hallucinations}. These works find that execution-based feedback repairs many failures but is bottlenecked by the specificity of the error signal. \system{}'s retry loop uses the same idea: observed accuracy gains are largest when the failure produces a detailed traceback rather than a generic safety-check rejection.

Each line of prior work above addresses part of the problem: log parsing recovers templates without consuming queries, LLM-based security analysis targets specific downstream tasks over already-parsed data, and general code- and query-generation systems either assume a schema or address non-security workloads. \system{} fills the gap by generating end-to-end query code from a natural-language security question against a raw, semi-structured log: without a pre-existing parser, without a pre-defined schema, and across queries that range from simple keyword filters to complex temporal correlations.

\section{\system{}}
\label{sec:system}

\system{} is a parser-free log-querying system that compiles a natural-language question into an executable extraction script and runs it deterministically against a raw log file. The user supplies a query and the path to a log; \system{} returns the matching records or extracted fields. Internally, an LLM is invoked once per query to emit either a self-contained Python script or a shell script (the language is a runtime choice), and the script is executed in a sandbox under a fixed set of safety, isolation, and output-format checks. The system is designed for ad hoc investigations and for the long tail of analyst queries that fall outside a SIEM's normalized schema; it is independent of any specific log source, schema, or detection language. We describe the pipeline and its inputs (\S\ref{sec:pipeline}) and the strategies for grounding the LLM in the source's format (\S\ref{sec:strategies}).

\subsection{Pipeline Overview}
\label{sec:pipeline}

Given a natural-language query $q$ and a security log file $L$, \system{} operates in four stages (Fig.~\ref{fig:system}).

\begin{enumerate}[leftmargin=*,label=\textbf{Stage \arabic*:},wide=0pt]
    \item \textbf{Prompt construction.}
    The system constructs a prompt from (i) the natural-language query $q$; (ii) context about the log, which can include templates $T$ and/or a subset $S$ of log lines drawn from $L$, depending on the chosen context strategy (\S\ref{sec:strategies}); (iii) constraints on the generated code, which must be Python or Bash and must not write files or open network sockets; and (iv) an output specification listing the column names and data types the generated code is expected to emit. Appendix~\ref{app:prompts} reproduces a complete prompt and the resulting generated code.

    \item \textbf{Code generation.}
    The assembled prompt is sent to an LLM. The response is parsed to extract either a self-contained Python script or a shell script. The model is instructed to return code in a JSON wrapper (\mbox{\texttt{\{"language":"python",}} \mbox{\texttt{"code":"..."\}}}), which allows reliable extraction even when the response contains explanatory text.

    \item \textbf{Sandboxed execution.}
    The generated code runs in a subprocess with the log file path as its sole argument. Execution is subject to three guardrails:

    \emph{Safety checks.} Before execution, the code is scanned for prohibited patterns. For Python, an AST walk rejects forbidden imports (e.g., \texttt{subprocess}), forbidden calls (e.g., \texttt{eval}, \texttt{exec}), any filesystem-write operation, and code that does not reference the target file via \texttt{sys.argv} or \texttt{argparse}. For Bash, a token scan rejects destructive operators (\texttt{rm}, \texttt{mv}, \texttt{sudo}, redirection). Code that fails the safety check is not executed; the rejection message is fed back to the model for retry.

    \emph{Execution isolation.} The subprocess runs with no outbound network access and no filesystem writes outside its working directory. A wall-clock timeout (1200 seconds in our evaluation) bounds execution time; queries that exceed this budget are terminated and scored as errors.

    \emph{Output-format contract.} After execution, the number of columns and their data types is checked against the specification in the prompt. Violations trigger a retry with the format error as feedback.

    \item \textbf{Retry and output.}
    When any stage fails, an error message is appended to the conversation history and the model is re-prompted. Up to $k{=}4$ retries are allowed. On success, we collect the script's standard output line by line: each line is one row of the answer (a matching log line for filter queries, or an extracted or aggregated record otherwise). We evaluate by comparing this answer set to the rule-based ground truth and computing set-based precision, recall, and $F_1$ (\S\ref{sec:groundtruth}).
\end{enumerate}

\paragraph{End-to-end example.}
For the query ``Find source IPs that produced 10 or more SSH authentication failures'', \system{} generates a Python script that uses \texttt{re} to match both \texttt{Failed password for ... from <ip>} and PAM-format \texttt{authentication failure; ... rhost=<ip>} lines, increments a \texttt{Counter} keyed by IP, and prints the IPs whose count meets the threshold. Running the script on the SSH log emits one row per offending IP with its count; that result set is the answer.

\subsection{Context Strategies}
\label{sec:strategies}

An LLM trained on public corpora has almost certainly seen examples of common security log formats, but it does not necessarily know every template, every field convention, or every quirk that a particular source of logs emits in production. The role of the context block in the prompt is to close this gap: to give the model just enough information about the source at hand for it to generate extraction code that is correct and that covers all relevant message formats. We evaluate five strategies that span the effort--information tradeoff.
\begin{enumerate}[leftmargin=*,label=\textbf{\arabic*.},wide=0pt]
    \item \textbf{Matryoshka templates.}
    We use the automated Matryoshka pipeline~\cite{piet2025matryoshka} to identify semantic templates with named field placeholders (e.g., \texttt{<SOURCE\_IP>}, \texttt{<AUTH\_METHOD>}). We then find one example log line per template. Matryoshka generates high quality templates, so this strategy measures how \system{} performs when given the highest effort templates available for these log sources.

    \item \textbf{Drain3.}
    The Drain3 streaming parser~\cite{he2017drain} extracts templates with generic \texttt{<*>} wildcards in a single pass over the log; we keep one example line per template (the first line in the log that produced it).

    \item \textbf{Frequency.}
    We developed a basic frequency-based templater inspired by IPLoM's token-count partitioning~\cite{makanju2009iploml}. The algorithm operates in four passes.
    \emph{(1) Pre-tokenization normalization.} Apply a small set of regular expressions to replace common high-variability sub-strings (IP addresses, hex values, quoted strings, deeply nested file paths) with a wildcard \texttt{<*>}. These regexes target syntactic shapes that recur across many log sources, rather than the schema of any particular source.
    \emph{(2) Tokenization and length partitioning.} Split each (now-normalized) message on whitespace and group messages by their token count.
    \emph{(3) Variable-position detection.} Within each length group $\ell$ containing $n_\ell$ messages, compute for each token position $p$ the unique-value ratio $r_p = u_p / n_\ell$, where $u_p$ is the number of distinct token strings appearing at position $p$. Mark $p$ as variable if $r_p$ exceeds a threshold (we use 0.3); otherwise it is constant. Replacing each variable position with \texttt{<*>} produces a per-line \emph{skeleton}, and lines sharing a skeleton collapse into one template.
    \emph{(4) Consolidation by structural signature.} Two skeletons with the same shape may differ only in their \texttt{key=value} tokens (e.g.\ \texttt{dev=<*>} vs.\ \texttt{ino=<*>} at the same position). For each skeleton we compute its \emph{structural signature}: the ordered sequence of non-wildcard tokens with each \texttt{key=value} reduced to \texttt{key=}. Skeletons sharing a signature collapse into one template, with the most frequent skeleton kept as the representative.
    After consolidation, we drop templates that occur fewer than $\max(2, n/10{,}000)$ times to remove one-off messages, and emit one example line per surviving template (the first line that produced it). The approach requires no external dependencies and no LLM calls, and runs in under ten seconds on every dataset in our evaluation.

    \item \textbf{Random sample.}
    The prompt includes 100 randomly sampled log lines but no templates, testing how much the model can infer about the log format from raw examples alone.

    \item \textbf{No context.}
    The prompt includes only a single sample line and no templates, testing the model's pre-trained knowledge of security log formats without any source-specific grounding.
\end{enumerate}

\section{Evaluation and Baselines}
\label{sec:eval}

This section describes our empirical setup: the five security log sources in our evaluation corpus (\S\ref{sec:datasets}), the two query tiers (\S\ref{sec:querydesign}), the rule-based ground truth (\S\ref{sec:groundtruth}), the precision/recall/$F_1$ metrics and variance protocol (\S\ref{sec:metrics}), and the human-analyst baseline (\S\ref{sec:human-baseline}).

\subsection{Security Log Datasets}
\label{sec:datasets}

We rely on the SecurityLogs corpus curated by Piet et~al.~\cite{piet2025matryoshka}, which extracts five security-relevant log sources from a Red Hat Bugzilla attachment crawl~\cite{redhatbugzilla} (Table~\ref{tab:datasets}). The five sources cover host-level access control (Linux Audit, including SELinux decisions and system-call traces), remote authentication and session activity (sshd), scheduled-job execution and a common persistence vector (cron), configuration management state changes whose failures can silently introduce vulnerabilities (Puppet), and dynamic address assignment relevant to network forensics (DHCP). Together they span 7 to 496 distinct templates per source, reflecting the format diversity that makes manual parser construction expensive in practice, and total 659{,}045 lines, the same corpus used to evaluate Matryoshka, enabling direct methodological comparison.

\begin{table}[t]
\caption{Dataset summary. simple = keyword filter, complex = multi-line correlation.}
\label{tab:datasets}
\centering\small
\begin{tabular}{lrS[table-format=2]S[table-format=2]}
\toprule
Log Type & Lines & {Simple Queries} & {Complex Queries} \\
\midrule
Audit (Linux) & 76,636 & 12 & 8 \\
SSH (sshd) & 35,329 & 12 & 8 \\
Cron & 12,547 & 12 & 8 \\
Puppet & 156,880 & 12 & 8 \\
DHCP & 377,653 & 32 & 21 \\
\midrule
\textbf{Total} & \textbf{659,045} & \textbf{80} & \textbf{53} \\
\bottomrule
\end{tabular}
\end{table}

\subsection{Query Design}
\label{sec:querydesign}

For each log type, our benchmark contains examples of all four query classes introduced in \S\ref{sec:queryclasses}: 80 queries in the simple tier (Classes 1--2) and 53 in the complex tier (Classes 3--4), for a total of \numqueries{} queries. The complex tier requires parsing timestamps, maintaining per-entity state across lines, and computing derived quantities, making it substantially harder for both manual scripting and LLM-generated code than the simple tier. The full text of every query appears in Appendix~\ref{app:queries}.

\subsection{Ground Truth Generation}
\label{sec:groundtruth}

For each log type, we implement a minimal hand-written Python parser that consumes the raw log line by line and emits structured records with typed fields (\texttt{line\_number}, \texttt{timestamp}, \texttt{host}, \texttt{process}, \texttt{message}, and event-specific fields such as \texttt{user}, \texttt{ip}, \texttt{port}). Each parser also exposes query-specific predicates that classify a record as an instance of a particular event type (e.g., an auth-failure predicate matches the union of PAM failures, failed-password events, and invalid-user events) and field extractors that pull specific values out of a record (e.g., a login extractor returns the user, IP, and port from a successful login). For each natural-language query, we then write a small Python procedure that uses these predicates and extractors to enumerate the ground-truth answer: matching raw lines for selection (\texttt{WHERE}) queries, extracted field tuples for projection (\texttt{SELECT}+\texttt{WHERE}) queries, and aggregated rows for grouped or correlation queries. The parsers and per-query procedures were verified manually against the raw logs through an iterative audit: we sampled lines from each source, and manually compared each predicate's classifications line by line against the natural-language intent. If necessary, we re-derived the aggregation logic for complex queries directly from the natural-language specification when manual inspection revealed a mismatch with the spec. All audits were conducted before experiments were run.

Certain WHERE queries are inherently ambiguous (e.g., ``return all network errors''). To address this, we categorize ground truth into three groups: definite matches (\texttt{must\_contain}), borderline/ambiguous matches (\texttt{may\_contain}), and definite non-matches. We use this distinction to define two scoring modes: \emph{strict} $F_1$ counts a returned row as a true positive only if it lies in \texttt{must\_contain}, and \emph{lenient} $F_1$ also accepts \texttt{may\_contain} matches. We report strict $F_1$ throughout the paper; lenient $F_1$ differs from strict by less than the run-to-run noise floor on every dataset and is reported per-query in the artifact.

\subsection{Evaluation Metrics}
\label{sec:metrics}

We evaluate the accuracy of \system{}-generated code by computing set-based precision, recall, and $F_1$ for each query. Macro $F_1$ is the unweighted average across queries within a dataset. To assess run-to-run stability, we repeat each condition $N{=}5$ times with a fixed sampling seed, so the only source of variance is LLM stochasticity. We report per-query standard deviations and pairwise significance tests in Appendix~\ref{app:variance}.

\subsection{Human Baseline}
\label{sec:human-baseline}

We manually write a script for each query, simulating a human analyst who spends a few minutes searching logs with standard tools.
An analyst investigating a raw log file typically reaches first for \texttt{grep} or \texttt{awk}. This works for filter queries but can struggle to express the operations complex queries demand: extracting structured fields, aggregating by entity, or joining records across lines. For select queries and complex multi-line queries that require state, the baseline is therefore a short Python script that uses \texttt{re} for field extraction and \texttt{Counter}/\texttt{defaultdict} for aggregation. All baseline scripts are released with the artifact (Appendix~\ref{app:open-science}) so that an external auditor can verify their representativeness.

\section{Results}
\label{sec:results}

All experiments use Gemini~3.1 Pro Preview~\cite{gemini2025} with Python output, 100 reservoir-sampled lines (where applicable), seed 42, and up to 4 retries (\S\ref{sec:pipeline}) unless otherwise noted; \S\ref{sec:language} compares Python and Bash, and \S\ref{sec:gpt-vs-gemini} repeats the benchmark with OpenAI GPT-5.4~\cite{gpt54systemcard} to test how the pipeline transfers across model families. Each subsection reports a separate set of LLM runs. \S\ref{sec:headline} reports the overall accuracy of \system{} under its cheapest practical configuration. \S\ref{sec:baselines} compares it against the two natural baselines: a hand-written human script (one-line \texttt{grep}/\texttt{awk} for filter queries, short Python for everything else) and an automated structured-parsing pipeline. \S\ref{sec:template-compare}--\S\ref{sec:gpt-vs-gemini} cover the design-space ablations: template strategy, prompt components, output language, and model family. Because code generation is stochastic, nominally identical configurations may produce slightly different $F_1$ scores across independent runs; our variance analysis (Appendix~\ref{app:variance}) quantifies this effect.

\subsection{Overall Accuracy}
\label{sec:headline}

Under \system{}'s cheapest practical configuration, the Drain3 templater~\cite{he2017drain} (which extracts log templates in a single linear pass over the input file with no LLM calls and finishes in under a minute on our largest dataset, DHCP at 378K lines), \system{} reaches macro $F_1$ 0.939 across the \numlogtypes{} log types (0.973 on simple queries, 0.905 on complex queries). The Drain3 column of Table~\ref{tab:template-compare} gives the per-cell breakdown. Switching to the more expensive Matryoshka context strategy (macro $F_1$ 0.947) lifts these numbers marginally, but the improvement is within run-to-run noise (\S\ref{sec:template-compare}) and does not justify the LLM cost of upfront parser synthesis.

In short, providing an LLM with lightweight log-format context is enough to generate query code that approaches the accuracy of structured parsing on simple queries and dominates manual scripting on complex ones, without an offline parser-generation step.

\subsection{Comparison to Baselines}
\label{sec:baselines}

\begin{table}[t]
\caption{Human baseline (\S\ref{sec:human-baseline}) vs.\ \system{}+Drain3, $F_1$ per dataset. \textbf{Bold} = winner per row.}
\label{tab:human}
\centering\small
\begin{tabular}{lrrr}
\toprule
Dataset & Human & \system{} & $\Delta$ \\
\midrule
Cron (S)   & \textbf{1.000} & 0.998 & $-$0.002 \\
SSH (S)    & \textbf{0.994} & 0.950 & $-$0.044 \\
Audit (S)  & \textbf{0.982} & 0.978 & $-$0.004 \\
DHCP (S)   & 0.703 & \textbf{0.989} & $+$0.286 \\
Puppet (S) & \textbf{0.999} & 0.951 & $-$0.048 \\
Cron (C)   & 0.906 & \textbf{0.991} & $+$0.085 \\
SSH (C)    & 0.742 & \textbf{0.847} & $+$0.105 \\
Audit (C)  & 0.659 & \textbf{0.910} & $+$0.251 \\
DHCP (C)   & 0.469 & \textbf{0.859} & $+$0.390 \\
Puppet (C) & 0.534 & \textbf{0.920} & $+$0.386 \\
\bottomrule
\end{tabular}
\end{table}

We compare \system{} against the two natural alternatives for log investigation: human-written scripts, which require no preprocessing but cost analyst effort at query time, and queries over Matryoshka-parsed structured logs, which make query writing cheap but require an expensive offline parsing step. \system{} approaches the accuracy of structured querying while outperforming manual scripting, combining the strengths of both: no preprocessing and low-effort queries.

\begin{figure}[t]
    \centering
    \includegraphics[width=\columnwidth]{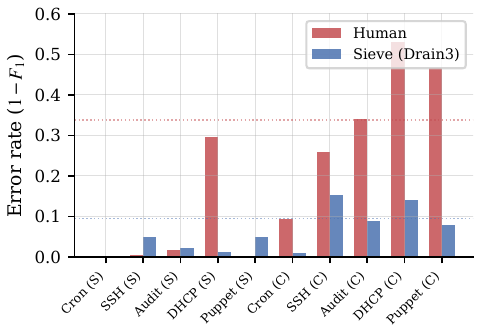}
    \caption{Error rate ($1 - F_1$) per dataset. \system{}+Drain3 reduces complex-tier error by 41--91\% vs.\ the human baseline.}
    \label{fig:errorrate}
\end{figure}

\noindent\textbf{Human baseline.} Table~\ref{tab:human} compares \system{} (with Drain3 context, same configuration as \S\ref{sec:headline}) against the human baseline defined in \S\ref{sec:human-baseline}: a one-line \texttt{grep}/\texttt{awk} for filter queries and a short Python script for select and complex queries. \system{} outperforms the human baseline on every complex dataset, by margins of $+$0.09 (Cron) to $+$0.39 (DHCP, Puppet). Framed as error rate ($1 - F_1$, which collapses missed results and false positives into a single measure), \system{} reduces the error rate by 3.6$\times$ (from 0.34 for the human baseline to 0.10 for \system{}; see Figure~\ref{fig:errorrate}). On simple queries \system{} performs similarly to the human baseline, except on DHCP ($+$0.29 for \system{}), where multiple message-format variants defeat any single grep pattern.

This performance reflects template coverage, not analyst skill or language choice. Many security queries hinge on enumerating every message-format variant in which a semantic event appears. A one-pass regex matches one or two variants at a time; the long tail of formats requires either prior knowledge of the source or a parser. Given the Drain3 template list, \system{} can branch over each variant in the generated code without the analyst having to recall or anticipate all of them. Two examples make this concrete:

\textit{DHCP simple, \texttt{dhcp\_query\_2}} (``return log lines reporting DHCP server IP addresses that are not broadcast addresses''). The human script is a one-line grep: \texttt{grep -v 255.255.255.255 | grep -E 'DHCP(OFFER|ACK)'}, which filters broadcast lines and matches the two DHCP message types where server IPs most often appear. The script captures only 42K of the 152K ground-truth lines (recall 0.28, $F_1$ 0.43), because non-broadcast server IPs also appear in DHCPREQUEST and DHCPRELEASE templates that the script does not enumerate. Adding more templates to the alternation requires the analyst to know in advance every message type that mentions a server IP. \system{}'s generated code reasons about all such templates from the Drain3 context and matches all 152K lines ($F_1$ 1.00).

\textit{Audit complex, \texttt{multiline\_4}} (``for each host and PAM action, count total PAM events and how many ended in failure''). The human script is a short Python program that extracts \texttt{op=PAM:<action>} from each line, groups by (host, action), and counts lines tagged \texttt{res=failed}. However, its single regex matches every PAM-related entry, including \texttt{pam\_unix} session-management messages that the ground truth treats as a different event class, and it returns 5{,}747 rows against a ground truth of 93 (precision 0.016). Distinguishing the relevant PAM templates from the irrelevant ones requires per-template branching against the audit subsystem's PAM event taxonomy, which the analyst would need to recall by hand but which the LLM derives from the Drain3 templates. \system{}'s generated code handles all relevant templates and reaches near-perfect precision and recall.

\begin{figure}[t]
    \centering
    \includegraphics[width=\columnwidth]{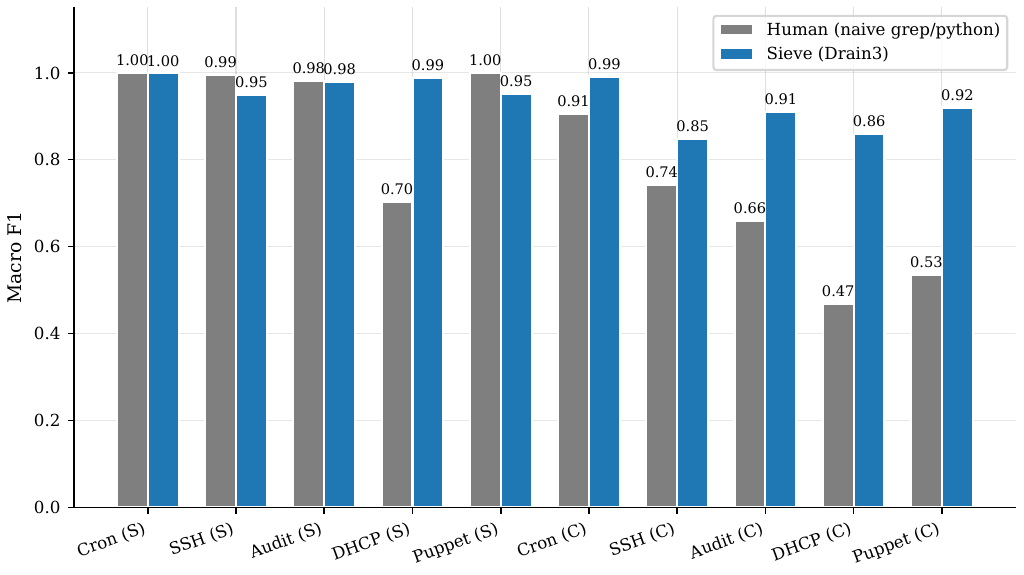}
    \caption{Human baseline vs.\ \system{}+Drain3, $F_1$ per dataset.}
    \label{fig:human}
\end{figure}

\begin{table}[t]
\caption{Matryoshka structured parsing vs.\ \system{}+Drain3 on a random sample of 5 simple \texttt{where} queries per log type (25 total).}
\label{tab:matry-vs-sieve}
\centering\small
\begin{tabular}{lrrr}
\toprule
Log type & Matryoshka $F_1$ & \system{} (Drain3) $F_1$ & $\Delta$ \\
\midrule
Audit  & 0.998 & 0.979 & $-$0.019 \\
Cron   & 1.000 & 1.000 & \phantom{$-$}0.000 \\
DHCP   & 1.000 & 1.000 & \phantom{$-$}0.000 \\
Puppet & 0.998 & 0.930 & $-$0.068 \\
SSH    & 0.999 & 0.995 & $-$0.004 \\
\midrule
Macro  & \textbf{0.999} & 0.981 & $-$0.018 \\
\bottomrule
\end{tabular}
\end{table}

\noindent\textbf{Structured-parsing baseline.} The other natural baseline is an automated parser-based pipeline. We use Matryoshka~\cite{piet2025matryoshka} as the reference: it generates per-source parsers via an LLM, applies them offline to materialize a structured representation, then runs queries over the parsed records. We sampled 5 simple \texttt{where} queries per log type (25 queries total), translated each natural-language query into a query over the Matryoshka-parsed representation, and scored the matched lines against the same \system{} ground truth. \system{} was run with Drain3.

Table~\ref{tab:matry-vs-sieve} reports per-log-type macro $F_1$. Matryoshka reaches 0.999 macro $F_1$ on the sample. \system{} reaches 0.981; on Cron, DHCP, and SSH the two pipelines are within 0.005 $F_1$, and on Audit \system{} is within 0.02. There is a larger gap for Puppet (0.07), where \system{}'s code for two queries over-match by a few extra lines (precision 0.60 and 0.82), reducing $F_1$ despite perfect recall. On simple filtering workloads, querying raw logs with LLM-generated code gets within noise of querying a fully parsed representation. 

\subsection{Design-Space Analysis}

We vary four design choices and measure their effect on accuracy: the source of log-format context, the prompt components, the generated language (Python vs.\ Bash), and the underlying LLM.

\subsubsection{Template Strategy Comparison}
\label{sec:template-compare}

\begin{table*}[t]
\begin{minipage}[t]{0.48\textwidth}
\centering
\captionof{table}{Macro $F_1$ across context strategies. S denotes simple, C denotes complex.}
\label{tab:template-compare}
\small
\begin{tabular}{lrrrrr}
\toprule
Dataset & Matry. & Drain3 & Freq. & Rand. & None \\
\midrule
Cron (S) & \textbf{0.998} & \textbf{0.998} & 0.917 & 0.750 & 0.915 \\
SSH (S) & 0.971 & 0.950 & 0.987 & \textbf{0.990} & 0.781 \\
Audit (S) & 0.984 & 0.978 & 0.978 & \textbf{0.990} & 0.680 \\
DHCP (S) & 0.945 & \textbf{0.989} & 0.919 & 0.891 & 0.916 \\
Puppet (S) & 0.945 & \textbf{0.951} & 0.875 & 0.856 & 0.772 \\
Cron (C) & 0.991 & 0.991 & \textbf{1.000} & 0.991 & 0.991 \\
SSH (C) & \textbf{0.892} & 0.847 & 0.846 & 0.863 & 0.601 \\
Audit (C) & \textbf{0.954} & 0.910 & 0.849 & 0.946 & 0.487 \\
DHCP (C) & \textbf{0.861} & 0.859 & 0.857 & 0.757 & 0.723 \\
Puppet (C) & 0.928 & 0.920 & \textbf{0.945} & 0.746 & 0.684 \\
\midrule
Average & \textbf{0.947} & 0.939 & 0.917 & 0.878 & 0.755 \\
\bottomrule
\end{tabular}
\end{minipage}
\hfill
\begin{minipage}[t]{0.48\textwidth}
\centering
\captionof{table}{Prompt ablation: $F_1$ when removing one component. S denotes simple, C denotes complex.}
\label{tab:ablation}
\small
\begin{tabular}{lrrrrr}
\toprule
Dataset & Full & $-$Tmpl & $-$Samp & $-$Retry & Query \\
\midrule
Cron (S) & \textbf{0.998} & 0.917 & \textbf{0.998} & \textbf{0.998} & 0.915 \\
SSH (S) & 0.971 & \textbf{0.988} & 0.986 & 0.903 & 0.934 \\
Audit (S) & 0.984 & \textbf{0.991} & 0.939 & 0.574 & 0.558 \\
DHCP (S) & 0.945 & 0.901 & \textbf{0.978} & 0.813 & 0.908 \\
Puppet (S) & \textbf{0.945} & 0.772 & 0.896 & 0.938 & 0.772 \\
Cron (C) & 0.991 & \textbf{0.998} & 0.995 & 0.741 & 0.995 \\
SSH (C) & \textbf{0.892} & 0.842 & 0.832 & 0.684 & 0.552 \\
Audit (C) & \textbf{0.954} & 0.795 & 0.946 & 0.479 & 0.307 \\
DHCP (C) & \textbf{0.861} & 0.773 & 0.765 & 0.617 & 0.731 \\
Puppet (C) & \textbf{0.928} & 0.517 & 0.819 & 0.490 & 0.597 \\
\midrule
Average & \textbf{0.947} & 0.849 & 0.915 & 0.724 & 0.727 \\
\bottomrule
\end{tabular}
\end{minipage}
\end{table*}

Table~\ref{tab:template-compare} compares five context strategies across all 10 datasets. The three template-backed strategies (Matryoshka, Drain3, Frequency) all substantially outperform the no-context baseline (macro $F_1$ 0.947, 0.939, 0.917 respectively, vs.\ 0.755 for ``none''), and they are statistically indistinguishable from one another (Appendix~\ref{app:variance}). Random samples alone (0.878) outperform no-context but trail template-backed strategies by 0.04--0.07 on average.

Template quality matters less than coverage. The model needs to see at least one example of each unique format the log produces; once that coverage is in place, the structural sophistication of the template adds little. Matryoshka contributes consistency, semantic field names, and a parser that can ingest logs at runtime. Drain3 splits common prefixes (e.g., one template per month abbreviation), producing narrower templates than Matryoshka but covering the same set of unique formats. The Frequency templater sits between the two: it consolidates skeletons by their structural signature, producing fewer templates than Drain3, but lacks Matryoshka's semantic field names. None of these differences materially change code quality on this benchmark: each strategy surfaces all unique formats, and the model writes the same kind of branching code regardless. What differs is cost. Frequency finishes in under ten seconds on every dataset; Drain3 takes seconds on smaller logs and about a minute on DHCP (378K lines). Both make zero LLM calls. Matryoshka's automated pipeline takes several hours and roughly \$0.60 in LLM fees per source~\cite{piet2025matryoshka}. The practical recommendation is to default to a lightweight templater.

When templates do help, it is because they expose structured fields that random samples may not surface. On sshd complex \texttt{multiline\_6} (``count failed SSH authentication events targeting the root account''), the no-context strategy scores 0.000 because the model writes a regex matching only OpenSSH's direct format (\texttt{Failed password for root from X}) and misses PAM-formatted entries that report the same authentication failure with a \texttt{user=root rhost=X} key-value structure. The grep ``root'' would match both, but the model's regex is anchored on the OpenSSH phrase. Template-backed strategies surface both formats and the model writes both branches, recovering the query.

The marginal value of templates is thus a function of log diversity rather than format sophistication. Cron has a narrow event vocabulary, and the model handles it without context (cron complex reaches 0.991 with ``none'' alone). Audit has rich \texttt{key=value} structure that is easy to misparse from raw lines (Audit~S ``none'' 0.680, Audit~C 0.487), and templates lift these to 0.98 and 0.95 respectively. DHCP, SSH, and Puppet sit in between: the model knows the format partially but template guidance helps mainly through coverage of message-type variants.

\subsubsection{Prompt Component Ablation}
\label{sec:ablation}

Table~\ref{tab:ablation} isolates the contribution of each prompt component (introduced in \S\ref{sec:pipeline}) by removing one element at a time; the \emph{Full} column reproduces the Matryoshka column of Table~\ref{tab:template-compare}. Removing templates costs 0.10 $F_1$ on average, with the largest hit on Puppet complex ($-$0.41) and Puppet simple ($-$0.17), where the format vocabulary is unfamiliar enough that the model cannot recover it from raw samples. Removing samples costs 0.03 $F_1$ on average and concentrates on DHCP complex ($-$0.10), where samples become the primary source of format information when the template is sparse. Removing retries costs 0.22 $F_1$, the largest single-component effect; removing all context (the rightmost ``query-only'' column) costs the same 0.22 on average but reaches $-$0.34 to $-$0.65 on complex datasets where a single sample line is no substitute for full context.

The retry loop catches three first-attempt failure modes: \emph{execution errors} where the script throws an exception, \emph{output-format errors} where the script runs but emits values in the wrong shape, and \emph{safety-check rejections} where the sandbox blocks code containing write primitives or subprocess calls. In the no-retries column, safety rejections alone account for 23\% of queries on average (range 0\% on cron and puppet simple to 50\% on audit complex), each scored $F_1{=}0$. With retries, the failure message is fed back to the model and the next attempt usually produces compliant code, consistent with prior work that finds execution-based feedback recovers shallow failures more reliably than semantic ones~\cite{chen2024selfdebug, olausson2024selfrepair}. Surface-level error feedback alone recovers 0.22 $F_1$ on average; a richer mechanism that also detects semantic deviations from the ground truth would likely help further on the hardest queries (\S\ref{sec:limitations}).

\begin{figure}[t]
    \centering
    \includegraphics[width=\columnwidth]{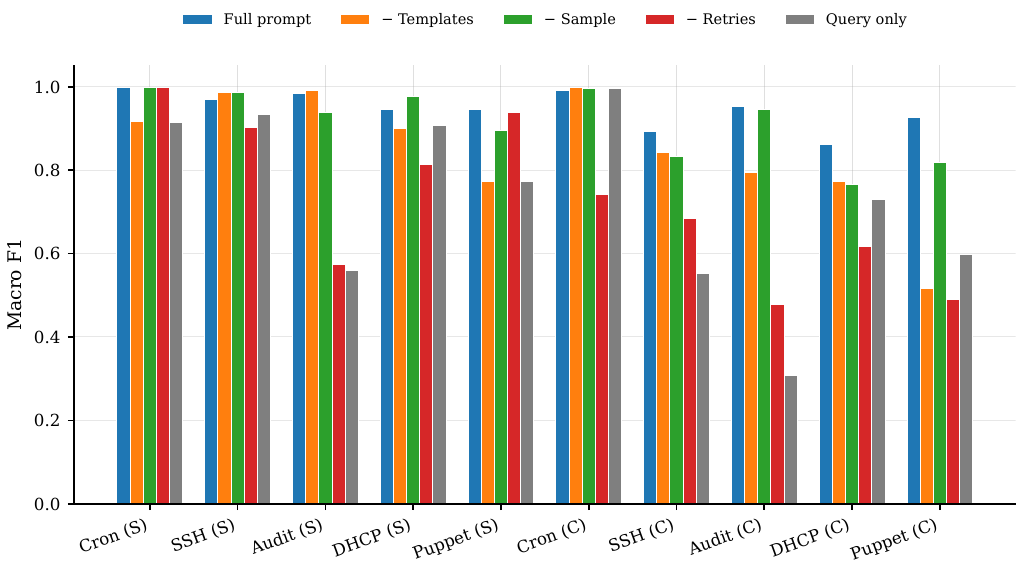}
    \caption{Prompt ablation: $F_1$ when each component of the prompt (\S\ref{sec:pipeline}) is removed. Retries contribute most; templates help most on specialized formats.}
    \label{fig:ablation}
\end{figure}

\subsubsection{Language Comparison}
\label{sec:language}

\begin{table}[t]
\caption{Language comparison: Bash (grep/awk) vs.\ Python.}
\label{tab:language}
\centering\small
\begin{tabular}{lrr}
\toprule
Dataset & Bash & Python \\
\midrule
Cron (S) & 0.915 & \textbf{0.998} \\
SSH (S) & \textbf{0.987} & \textbf{0.987} \\
Audit (S) & \textbf{0.989} & \textbf{0.989} \\
DHCP (S) & \textbf{0.978} & 0.960 \\
Puppet (S) & \textbf{0.998} & \textbf{0.998} \\
Cron (C) & \textbf{0.993} & 0.991 \\
SSH (C) & 0.904 & \textbf{0.938} \\
Audit (C) & 0.574 & \textbf{0.929} \\
DHCP (C) & 0.446 & \textbf{0.815} \\
Puppet (C) & 0.773 & \textbf{0.861} \\
\bottomrule
\end{tabular}
\end{table}

Table~\ref{tab:language} compares Bash and Python as the generated language. On simple queries, Bash and Python differ by less than 0.02 $F_1$ on four of five datasets, with cron simple as the exception (Bash 0.915 vs Python 0.998). Both languages are adequate for keyword filters (\texttt{grep -E 'pattern'} and \texttt{line.find(pattern)} are functionally equivalent for single-line matching), and language choice is not a meaningful design variable on the simple tier in general.

On complex datasets the languages diverge: Python averages 0.17 $F_1$ above Bash, with the gap concentrated on datasets that demand stateful temporal reasoning. The DHCP complex gap is the largest: Bash achieves 0.446 while Python achieves 0.815, a difference of 0.37 $F_1$. Of 21 DHCP complex queries, Bash produces $F_1{=}0$ on 10, whereas Python does so on only 2. The failures follow a consistent pattern: queries requiring per-transaction joins, temporal windows, or lease-cycle tracking demand data structures (dictionaries keyed by transaction ID, timestamp arithmetic across day boundaries, per-entity state machines) that Python's standard library provides idiomatically but that Bash solutions must approximate through \texttt{awk} associative arrays and string-based timestamp comparisons. These approximations break on edge cases (day-boundary rollovers, multi-field composite keys, nested groupings), producing complete failures on the queries where LLM-generated code provides the greatest advantage. On DHCP queries that reduce to set-membership or counting operations (\texttt{grep + sort + uniq} patterns) by their natural-language specification, Bash matches Python, confirming the language gap is specific to stateful computation.

The tradeoff is asymmetric. Python's downside on simple queries is slightly more verbose code; Bash's downside on complex queries is outright failure. \emph{Python should therefore be the default output language for LLM-generated log query code}, with Bash reserved for cases where an analyst explicitly requests a one-liner for a simple filter.

\subsubsection{Model Comparison}
\label{sec:gpt-vs-gemini}

\begin{table}[t]
\caption{\system{}+Matryoshka under Gemini~3.1 Pro Preview vs.\ GPT-5.4.}
\label{tab:gpt-vs-gemini}
\centering\small
\begin{tabular}{lrrr}
\toprule
Dataset & Gemini~3.1 Pro & GPT-5.4 & $\Delta$ \\
\midrule
Audit (S)  & 0.984 & 0.913 & $-$0.071 \\
Audit (C)  & 0.954 & 0.846 & $-$0.108 \\
Cron (S)   & 0.998 & 0.998 & \phantom{$-$}0.000 \\
Cron (C)   & 0.991 & 0.995 & $+$0.004 \\
DHCP (S)   & 0.944 & 0.975 & $+$0.031 \\
DHCP (C)   & 0.861 & 0.467 & $-$0.394 \\
Puppet (S) & 0.945 & 0.914 & $-$0.031 \\
Puppet (C) & 0.927 & 0.638 & $-$0.289 \\
SSH (S)    & 0.971 & 0.766 & $-$0.205 \\
SSH (C)    & 0.892 & 0.871 & $-$0.021 \\
\midrule
Simple avg  & 0.968 & 0.913 & $-$0.055 \\
Complex avg & 0.925 & 0.764 & $-$0.161 \\
Macro avg   & \textbf{0.947} & 0.838 & $-$0.108 \\
\bottomrule
\end{tabular}
\end{table}

To test how much of \system{}'s accuracy is tied to a specific model family, we ran the full benchmark using OpenAI's GPT-5.4~\cite{gpt54systemcard} in place of Gemini~3.1 Pro Preview, holding the prompt, sample size, retry policy, and Matryoshka context strategy fixed. Table~\ref{tab:gpt-vs-gemini} reports per-dataset macro $F_1$ for both models.

GPT-5.4 reaches 0.838 macro $F_1$ versus Gemini's 0.947, a gap of 0.108. On Cron the two models are at parity (within 0.01), and on DHCP simple GPT-5.4 in fact leads by 0.03. The gap concentrates on three datasets: DHCP complex ($-$0.39), Puppet complex ($-$0.29), and SSH simple ($-$0.21). DHCP complex is the largest: 7 of its 21 queries return $F_1{=}0$ because GPT-5.4's generated code does not correctly track per-transaction state across REQUEST/ACK pairs or join clients to MAC and IP across line variants. The pipeline itself is vendor-neutral: across all 133 queries, zero hit safety-check rejections, retries, or API errors after the OpenAI shim landed; every $F_1{=}0$ case is a script that ran cleanly to completion and returned the wrong rows.

The size of the gap on our task is not predicted by general rankings. On standard public benchmarks the two models are close, with GPT-5.4 leading Gemini~3.1 Pro on SWE-bench Verified~\cite{jimenez2024swebench} (71.7\% vs 63.8\%) and the two tied on the Artificial Analysis Intelligence Index. Two non-mutually-exclusive explanations are plausible. First, task specificity: our workloads exercise per-entity state machines, key-value parsing under multiple template variants, and rolling-window aggregation, none of which feature prominently in standard coding benchmarks. Second, prompt-tuning bias: we developed the prompt on a small subset of DHCP queries using Gemini and did not modify it after that, so the prompt may be unintentionally specialized to Gemini's instruction-following style.

\system{}'s pipeline transfers across model families without modification, and on four of five simple datasets the two models are within 0.07 $F_1$ (with GPT-5.4 ahead on DHCP simple). On the complex tier and on SSH simple GPT-5.4 trails, but disentangling how much of that gap is model capability and how much is prompting quirk would require either re-tuning the prompt against GPT-5.4 or running each prompt variant against both models in a controlled multi-run study; both are future work.

\subsection{Failure Mode Analysis}
\label{sec:failures}

We manually inspected the generated code for every query where the Matryoshka strategy scored strict $F_1 < 0.5$. The dominant failure mode is incomplete rule enumeration on stateful queries: the model handles the dominant cases correctly but misses a subset of the rules, even when the natural-language specification is unambiguous.

DHCP complex \texttt{multiline\_13} (``identify invalid DHCP cycle transitions'') is an instance. Per the query, after a \texttt{DHCPACK} the transaction is complete, and any further message sharing the same transaction ID is a violation. The generated code's list of allowed message-type transitions incorrectly includes \texttt{DHCPACK} followed by \texttt{DHCPDISCOVER} (and a few other after-\texttt{ACK} pairs), so any violation that hinges on a message appearing after the closing \texttt{ACK} gets classified as clean. \system{} correctly flags the violations its list does cover (precision 0.998) but misses most of the rest (recall 0.19).

The general pattern is that when a query requires enumerating many rules, the model sometimes captures the high-frequency cases and overlooks the remainder, though this is certainly not always the case. We discuss the broader implications for code generation systems in \S\ref{sec:limitations}.

\section{Discussion}
\label{sec:discussion}

\subsection{When parser-free querying helps}
\label{sec:when-helps}

\system{} and parser-based ingestion address different operational regimes. Static parsers are the right tool when a small number of high-volume sources drive a fixed set of standing detection rules: ingestion is regex matching, and queries can be handled efficiently using database indices. \system{} is the right tool for the long tail of queries that the parser author did not anticipate: log message formats that aren't reflected in the parser, log sources for which no parser exists, and ad hoc investigations under time pressure. On queries that require per-entity state, structured-field extraction, or temporal joins, \system{}'s per-query LLM cost (median 63 seconds) buys a substantial accuracy improvement over a quick manual script. A production deployment can run both paths: parser-generation pipelines such as Matryoshka for high-volume standing rules, and \system{} for the long tail of analyst queries.

\subsection{Deployment}
\label{sec:deployment}

Our evaluation runs against static log files; in practice logs are live streams. A natural architecture might pair a streaming template miner (e.g., Drain~\cite{he2017drain}, which processes each line in constant time against a fixed-depth prefix tree) with query-time code generation. The miner runs continuously on the ingestion path and maintains a per-source index of (template, representative lines) pairs; its overhead is modest (Drain processes our largest dataset, DHCP at 378K lines, in about a minute, and its memory scales with the number of distinct templates rather than with total line count). At query time the system retrieves the current template index for the relevant source, assembles the prompt, generates code, and executes the script against the analyst's log slice. Format drift is absorbed incrementally as new templates appear, without offline regeneration. Within this architecture our results recommend Python as the default code-generation target (Python preserves accuracy on stateful queries where Bash collapses, \S\ref{sec:language}) and a lightweight templater (Drain3 or our Frequency variant) as sufficient grounding (Appendix~\ref{app:variance}); the per-query LLM cost is small enough that code generation is a good default for any query.

\subsection{Limitations}
\label{sec:limitations}

\paragraph{Model and dataset scope.}
We evaluate two model families (Gemini~3.1 Pro Preview and OpenAI GPT-5.4~\cite{gpt54systemcard}, \S\ref{sec:gpt-vs-gemini}), and Appendix~\ref{sec:model-compare} adds Gemini~3 Flash Preview~\cite{gemini3flash2026} as preliminary evidence that the qualitative findings transfer across model sizes within a vendor. Broader validation across additional vendors and a controlled multi-run study to disentangle model-quality differences from single-shot variance remain future work. The \numlogtypes{} log types in our benchmark come from a single source corpus (Red Hat Bugzilla); enterprise environments may exhibit different format characteristics, and the cron dataset in particular has limited template diversity.

\paragraph{Ground truth and query ambiguity.}
Ground truth is authored by us. We mitigate potential bias with dual-level scoring (strict \texttt{must\_contain} and lenient \texttt{may\_contain}) and by releasing all parsers and human baseline scripts for external audit. Natural-language queries also carry inherent ambiguity: different reasonable interpretations can yield different ground-truth sets. We removed ambiguous wording during dataset construction where possible, but the issue is fundamental to any natural-language interface over heterogeneous data.

\paragraph{Semantic correctness and retries.}
Successful execution does not guarantee semantic correctness: a small fraction of queries oscillate between $F_1{=}0$ and $F_1{=}1$ across runs (Appendix~\ref{app:variance}). Our retry loop catches only surface-level execution errors, yet even that signal recovers 0.22 $F_1$ on average (\S\ref{sec:ablation}). A natural direction for future work is a richer iterative mechanism that detects and corrects semantic deviations rather than syntactic faults alone; the size of the surface-level retry lift suggests a semantic-aware loop could close additional ground.

\paragraph{Multi-log queries.}
Our current pipeline targets a single log at a time. A natural extension is queries that span multiple related log files, for example correlating sshd authentication failures with audit subsystem events, or DHCP lease changes with downstream Puppet failures. The parser-free design extends naturally to a small set of co-sampled sources.

\system{}'s exposure to adversarial log content (prompt injection, information leakage) is a deployment-context concern rather than a blocking limitation. The intended use case is interactive analyst querying on logs that have already been collected and validated by the SOC's existing pipeline; in this regime, an attacker-crafted log line that could mislead the LLM is content the analyst would already need to reason about. We discuss the threat model and the sandbox-level mitigations in detail in Appendix~\ref{app:security}.

\section{Conclusion}
\label{sec:conclusion}

We presented \system{}, a parser-free system that compiles natural-language security questions into executable extraction code over raw logs. Across \numqueries{} queries on \numlogtypes{} log sources, \system{} matches or exceeds manual analyst scripting on complex temporal and cross-event queries, and a lightweight LLM-free templater is sufficient grounding in place of more expensive LLM-driven alternatives.

\begin{acks}
This research was supported by the KACST-UCB Joint Center on Cybersecurity, OpenAI, the National Science Foundation under grant IIS-2229876 (the ACTION center), Open Philanthropy, Google, and the Noyce Foundation. Any opinions, findings, and conclusions or recommendations expressed in this material are those of the author(s) and do not necessarily reflect the views of the sponsors.
\end{acks}

\bibliographystyle{ACM-Reference-Format}
\bibliography{references}

\appendix

\section{Ethical Considerations}
\label{app:ethics}

\paragraph{Stakeholders.}
Our work potentially impacts multiple groups: (i) security analysts and organizations that rely on log data for detection and forensics, (ii) end users whose activities are indirectly reflected in log data, (iii) service providers that generate and process logs, and (iv) the research community developing AI-driven security analytics.

\paragraph{Impacts and Principles.}
Following the Menlo Report~\cite{menlo2012}, we considered:
\begin{itemize}
    \item \emph{Beneficence.} Our goal is to improve security monitoring efficiency by reducing the manual effort required to author per-source parsers and per-query extraction code.
    \item \emph{Respect for Persons.} We only used public datasets (the SecurityLogs corpus curated from Red Hat Bugzilla attachments~\cite{piet2025matryoshka, redhatbugzilla}); no enterprise data was processed.
    \item \emph{Justice.} Results, code, queries, and ground-truth parsers are shared via an open release to ensure broad benefit.
    \item \emph{Respect for Law and Public Interest.} We adhered to the source dataset's data policies and avoided any live system interaction.
\end{itemize}

\paragraph{Potential Harms and Mitigations.}
\begin{itemize}
    \item \emph{Privacy.} We processed only public log data and share only aggregate, vetted results; the raw logs are referenced through the original sources rather than redistributed.
    \item \emph{Operational.} All experiments were run offline against static log files to avoid any production impact, and the generated code executes in a sandbox that blocks outbound network access, filesystem writes, and subprocess spawning (\S\ref{sec:pipeline}).
    \item \emph{Dual use.} An attacker with access to \system{} could in principle query stolen log data more efficiently, but the same capability is already available through manual scripting and existing SIEM tools, so \system{} does not introduce a qualitatively new risk.
\end{itemize}

We judged the benefits (faster, more accurate log querying for defenders and a reproducible benchmark) to outweigh residual risks.

\section{Open Science}
\label{app:open-science}

Our source code, prompts, the \numqueries{}-query benchmark with rule-based ground truth, the hand-written ground-truth parsers, the human baseline scripts, the evaluation framework, and the per-eval results for every experiment reported in this paper are available at \url{https://github.com/dr4Nx/sieve-public}.

\section{Generative AI Usage}
\label{app:genai}

Large language models constitute a core component of the system under evaluation: \system{} dispatches one model invocation per query to compile the analyst's natural-language question into executable extraction code, and our experimental results characterize the behavior of these models within that pipeline. The primary model used in the evaluation is Gemini~3.1 Pro Preview~\cite{gemini2025}; OpenAI GPT-5.4~\cite{gpt54systemcard} is examined as a second reference point in \S\ref{sec:gpt-vs-gemini}, and Gemini~3 Flash Preview~\cite{gemini3flash2026} is included as a third reference point reported in Appendix~\ref{sec:model-compare}.

Second, the authors employed generative AI tools to assist with the drafting and revision of this manuscript. All AI-assisted text was reviewed, edited, and approved by the authors, who bear sole responsibility for the paper's accuracy and presentation. Citations and quantitative claims were verified against their sources.

\section{Security Considerations}
\label{app:security}

\system{} executes LLM-generated code over log data that may contain attacker-influenced content. We discuss the threat model and the corresponding mitigations here.

\paragraph{Threat model.}
\system{} is intended for interactive analyst querying of logs that have already been collected by the SOC's existing pipeline. In this regime, the analyst already has read access to the logs in question, and any attacker-crafted line that could influence the LLM is also content the analyst could read (and would typically need to reason about) directly. We do not target online ingestion-time analysis on untrusted streams, which is the regime better served by parser-based ingestion such as Matryoshka~\cite{piet2025matryoshka}.

\paragraph{Prompt injection.}
A malicious actor who can write entries to a monitored source could craft lines designed to influence the model when those lines appear in the sampled prompt context. The prompt instructs the model to treat sample lines as structural examples for inferring log format rather than as commands. The output-format check (\S\ref{sec:pipeline}) and sandbox restrictions further constrain the consequences: even if the model produces non-target code, the sandbox blocks network access, filesystem writes, and subprocess spawning. These are defense-in-depth measures, not guarantees: a sufficiently crafted injection could in principle steer the model toward incorrect but plausible-looking extraction code that passes safety checks but returns wrong results, which is the same failure mode discussed in §\ref{sec:limitations} for benign queries.

\paragraph{Information leakage.}
The generated code reads the source log in its entirety and writes results to stdout. Outbound network access is blocked, so exfiltration via generated code is bounded to whatever appears in the analyst-visible output. In any production deployment, that output should be treated with the same sensitivity classification as the source log.

\paragraph{Higher-assurance deployments.}
For deployments that include partly untrusted log sources, the sandbox can be layered with additional controls: input-redaction passes that strip suspected attacker-controlled content before sampling, mandatory human review of generated scripts before execution against critical sources, and repeated generation with output cross-checking to flag semantically inconsistent results between runs.

\section{Evaluation Queries}
\label{app:queries}
All \numqueries{} ground-truth queries are listed below, grouped by log type and complexity tier. Each row gives the query identifier, the query type (\texttt{where} for line-level filters, \texttt{select} for field extraction and aggregation), the natural-language prompt sent to the model, and the expected output format.


\begin{table*}[!htbp]
\caption{Audit log queries (20 total: simple and complex).}
\label{tab:queries-audit}
\centering\scriptsize
\renewcommand{\arraystretch}{1.15}
\begin{tabular}{@{}p{1.6cm}ccp{11.5cm}@{}}
\toprule
\textbf{ID} & \textbf{Tier} & \textbf{Type} & \textbf{Natural-language query} \\
\midrule
\texttt{audit\_query\_1} & S & W & Find audit log lines that report SELinux AVC denials. \\
\texttt{audit\_query\_2} & S & W & Find audit log lines that show SELinux policy loads. \\
\texttt{audit\_query\_3} & S & W & Find audit log lines that show SELinux boolean changes. \\
\texttt{audit\_query\_4} & S & W & Find audit log lines that show SELinux enforcing mode changes. \\
\texttt{audit\_query\_5} & S & W & Find audit log lines showing kernel audit initialization events. \\
\texttt{audit\_query\_6} & S & W & Find audit log lines that record the audit daemon process ID being set or changed. \\
\texttt{audit\_query\_7} & S & W & Find audit log lines for failed SSH authentication attempts handled by sshd. \\
\texttt{audit\_query\_8} & S & W & Find audit log lines for PAM session open events. \\
\texttt{audit\_query\_9} & S & W & Find audit log lines for systemd unit start events that name a service unit. \\
\texttt{audit\_select\_1} & S & S & Find the hosts that logged SELinux AVC denials. \\
\texttt{audit\_select\_2} & S & S & Find the SELinux boolean names that were changed. \\
\texttt{audit\_select\_3} & S & S & Find the service unit names reported in systemd unit start events. \\
\midrule
\texttt{multiline\_1} & C & S & For each host, count SELinux AVC denial events and the number of distinct denied permissions. \\
\texttt{multiline\_2} & C & S & For each host, count SELinux policy load events and enforcing mode changes. \\
\texttt{multiline\_3} & C & S & Identify hosts and SELinux booleans where the same boolean was changed from enabled to disabled and later changed back to enabled. Return host, boolean\_name, disable\_event\_timestamp, and enable\_event\_timestamp. \\
\texttt{multiline\_4} & C & S & For each host and PAM action, count total PAM events and how many of those events ended in failure. \\
\texttt{multiline\_5} & C & S & For each source address, count failed SSH authentication attempts recorded by sshd. \\
\texttt{multiline\_6} & C & S & For each systemd service unit, count unit start events and unit stop events. \\
\texttt{multiline\_7} & C & S & For each host, count audit subsystem initialization events and audit daemon process ID set events. \\
\texttt{multiline\_8} & C & S & For each executable path and signal number, count crash signal events. \\
\bottomrule
\end{tabular}
\end{table*}

\begin{table*}[!htbp]
\caption{sshd log queries (20 total: simple and complex).}
\label{tab:queries-sshd}
\centering\scriptsize
\renewcommand{\arraystretch}{1.15}
\begin{tabular}{@{}p{1.6cm}ccp{11.5cm}@{}}
\toprule
\textbf{ID} & \textbf{Tier} & \textbf{Type} & \textbf{Natural-language query} \\
\midrule
\texttt{sshd\_query\_1} & S & W & Find log lines recording SSH authentication failures, including Failed password, Invalid user, and PAM authentication failure events. \\
\texttt{sshd\_query\_2} & S & W & Find log lines showing successful SSH logins using password authentication. \\
\texttt{sshd\_query\_3} & S & W & Find log lines showing successful SSH logins using public key authentication. \\
\texttt{sshd\_query\_4} & S & W & Find log lines where an SSH login was attempted with an invalid or unknown username. \\
\texttt{sshd\_query\_5} & S & W & Find log lines showing SSH session openings. \\
\texttt{sshd\_query\_6} & S & W & Find log lines showing SSH session closings. \\
\texttt{sshd\_query\_7} & S & W & Find log lines showing SSH network-level disconnection events. \\
\texttt{sshd\_query\_8} & S & W & Find log lines showing the SSH server starting and listening on a port. \\
\texttt{sshd\_query\_9} & S & W & Find log lines reporting deprecated SSH configuration options. \\
\texttt{sshd\_select\_1} & S & S & Find the source IP addresses or hostnames that caused SSH authentication failures (including Failed password, Invalid user, and unknown user attempts). \\
\texttt{sshd\_select\_2} & S & S & Find the usernames that were targeted in SSH authentication failure attempts (including Failed password, Invalid user, and unknown user attempts). \\
\texttt{sshd\_select\_3} & S & S & Find the ports on which the SSH server was configured to listen. \\
\midrule
\texttt{multiline\_1} & C & S & For each source IP that attempted SSH authentication and failed one or more times, count the total number of failed authentication events and the number of distinct usernames tried. Exclude connection-level events (port scans, protocol probes, connection resets) where no authentication was attempted. \\
\texttt{multiline\_2} & C & S & Find source IPs that attempted SSH authentication and failed 10 or more times within any 5-minute window. Return the source IP and the peak failure count within a single 5-minute window. Exclude connection-level events where no authentication was attempted. \\
\texttt{multiline\_3} & C & S & Find source IPs that attempted SSH authentication and failed using 5 or more distinct usernames. Return the source IP and the number of distinct usernames tried. Exclude connection-level events where no authentication was attempted. \\
\texttt{multiline\_4} & C & S & Find users who had successful SSH logins from two different source IP addresses within 60 seconds of each other. Assume year 2000 for timestamps. Return the username and the two IP addresses. \\
\texttt{multiline\_5} & C & S & For each host, count SSH session opens and session closes, and compute the number of unclosed sessions (opens minus closes). \\
\texttt{multiline\_6} & C & S & For each source IP, count the number of failed SSH authentication events that specifically targeted the root account. Exclude connection-level events where no authentication was attempted. \\
\texttt{multiline\_7} & C & S & For each host and username with successful SSH logins, count the total logins and the number of distinct key fingerprints used. Only count 'Accepted' login events, not session events. \\
\texttt{multiline\_8} & C & S & For each host, count the number of distinct source IPs that attempted SSH authentication and failed, and the number of distinct source IPs with successful SSH logins. Exclude connection-level events (port scans, protocol probes, connection resets) where no authentication was attempted. \\
\bottomrule
\end{tabular}
\end{table*}

\begin{table*}[!htbp]
\caption{cron log queries (20 total: simple and complex).}
\label{tab:queries-cron}
\centering\scriptsize
\renewcommand{\arraystretch}{1.15}
\begin{tabular}{@{}p{1.6cm}ccp{11.5cm}@{}}
\toprule
\textbf{ID} & \textbf{Tier} & \textbf{Type} & \textbf{Natural-language query} \\
\midrule
\texttt{cron\_query\_1} & S & W & Find log lines showing cron job command executions. \\
\texttt{cron\_query\_2} & S & W & Find log lines showing cron session openings. \\
\texttt{cron\_query\_3} & S & W & Find log lines showing cron session closings. \\
\texttt{cron\_query\_4} & S & W & Find log lines showing cron sessions opened for the root user. \\
\texttt{cron\_query\_5} & S & W & Find cron command-execution log lines on July 14, 2017. \\
\texttt{cron\_query\_6} & S & W & Find log lines mentioning the RANDOM\_DELAY scaling factor. \\
\texttt{cron\_query\_7} & S & W & Find log lines showing that the cron daemon started with inotify support. \\
\texttt{cron\_query\_8} & S & W & Find all cron log lines from July 14, 2017 between 03:00 and 04:00. \\
\texttt{cron\_query\_9} & S & W & Find all cron log lines with process ID 21832. \\
\texttt{cron\_select\_1} & S & S & Find the distinct commands that were executed by cron jobs. \\
\texttt{cron\_select\_2} & S & S & Find the usernames under which cron jobs were executed. \\
\texttt{cron\_select\_3} & S & S & Find the distinct dates on which cron activity was logged. \\
\midrule
\texttt{multiline\_1} & C & S & For each distinct command executed by cron, count how many times it was executed and by how many distinct users. \\
\texttt{multiline\_2} & C & S & For each hour of the day (0 through 23), count the total number of cron job command executions. \\
\texttt{multiline\_3} & C & S & For each date, count the number of cron session opens, session closes, and command executions. \\
\texttt{multiline\_4} & C & S & For each cron command, find the maximum elapsed time in seconds between any two consecutive executions of that same command. Floor the result to the nearest whole second (e.g. 900.99s becomes 900). Return the command and the maximum gap. This identifies missed or delayed scheduled executions. \\
\texttt{multiline\_5} & C & S & Find cron sessions (by host and process ID) where a session was opened but no command was executed in the same process. \\
\texttt{multiline\_6} & C & S & For each distinct command executed by cron, return the earliest and latest execution timestamp text. \\
\texttt{multiline\_7} & C & S & For each date and 15-minute time interval, count the number of cron command executions. The interval is represented as minutes since midnight (0, 15, 30, 45, 60, ...). Return only intervals that had at least one execution. \\
\texttt{multiline\_8} & C & S & For each host and date, count the number of cron command executions and session opens. \\
\bottomrule
\end{tabular}
\end{table*}

\begin{table*}[!htbp]
\caption{Puppet log queries (20 total: simple and complex).}
\label{tab:queries-puppet}
\centering\scriptsize
\renewcommand{\arraystretch}{1.15}
\begin{tabular}{@{}p{1.6cm}ccp{11.5cm}@{}}
\toprule
\textbf{ID} & \textbf{Tier} & \textbf{Type} & \textbf{Natural-language query} \\
\midrule
\texttt{puppet\_query\_1} & S & W & Find log lines indicating Puppet runs were disabled by an administrative action. \\
\texttt{puppet\_query\_2} & S & W & Find log lines indicating Puppet execution was re-enabled after being disabled. \\
\texttt{puppet\_query\_3} & S & W & Find log lines where Puppet says it could not fetch the node definition it needed. \\
\texttt{puppet\_query\_4} & S & W & Find log lines showing the agent could not retrieve its catalog from a remote source. \\
\texttt{puppet\_query\_5} & S & W & Find log lines showing the agent fell back to a cached catalog after a remote problem. \\
\texttt{puppet\_query\_6} & S & W & Find log lines showing Puppet could not deliver its report back to a remote service. \\
\texttt{puppet\_query\_7} & S & W & Find log lines showing resources were skipped because a dependency chain had already failed. \\
\texttt{puppet\_query\_8} & S & W & Find log lines where Puppet reports that a needed command could not be found. \\
\texttt{puppet\_query\_9} & S & W & Find log lines showing certificate validation or trust failures during remote Puppet communication. \\
\texttt{puppet\_select\_1} & S & S & Find the hosts that experienced remote catalog retrieval failures. \\
\texttt{puppet\_select\_2} & S & S & Find the configuration versions that were applied during Puppet runs. \\
\texttt{puppet\_select\_3} & S & S & Find the resource identifiers that were skipped because of failed dependencies. \\
\midrule
\texttt{multiline\_1} & C & S & For each host, count Puppet runs that begin applying a configuration version, count how many of those runs later finish, and return the completion rate. \\
\texttt{multiline\_2} & C & S & For each host, average the runtime reported for finished catalog runs and return the host with the average runtime in seconds, rounded to two decimal places. \\
\texttt{multiline\_3} & C & S & Find Puppet agent processes where a node-definition fetch failure is later followed by use of a cached catalog. Return the host, process id, and the original syslog timestamp text from those two events. \\
\texttt{multiline\_4} & C & S & For each host that saw report-send failures or remote catalog retrieval failures, count both and return those two totals. \\
\texttt{multiline\_5} & C & S & For each Puppet module name appearing after '/Stage[main]/' in skipped resource identifiers, count how many resources were skipped because of failed dependencies. \\
\texttt{multiline\_6} & C & S & For each host, group Puppet failure messages into these error families when they apply: certificate\_verify\_failed, getaddrinfo, network\_unreachable, command\_not\_found, timeout, error\_downloading\_packages, and no\_child\_processes. Return the host, the error family name, and the count. \\
\texttt{multiline\_7} & C & S & For each host and Puppet agent process that logged refresh activity, count how many refreshes were scheduled and how many refreshes were triggered. \\
\texttt{multiline\_8} & C & S & Find Puppet agent processes where a remote catalog retrieval failure is followed by use of a cached catalog and then a report-send failure. Return the host, process id, and the original syslog timestamp text from those three events. \\
\bottomrule
\end{tabular}
\end{table*}

\begin{table*}[!htbp]
\caption{DHCP log queries (53 total: simple and complex).}
\label{tab:queries-dhcp}
\centering\scriptsize
\renewcommand{\arraystretch}{1.15}
\begin{tabular}{@{}p{1.6cm}ccp{11.5cm}@{}}
\toprule
\textbf{ID} & \textbf{Tier} & \textbf{Type} & \textbf{Natural-language query} \\
\midrule
\texttt{dhcp\_query\_1} & S & W & Return lines assigning an IP to host laphroaig \\
\texttt{dhcp\_query\_2} & S & W & Return log lines reporting DHCP server IP addresses that are not broadcast addresses \\
\texttt{dhcp\_query\_3} & S & W & Return all log lines for transaction ID 0x6520bf0e \\
\texttt{dhcp\_query\_4} & S & W & Get all lines reporting MAC addresses for host laphroaig \\
\texttt{dhcp\_query\_5} & S & W & List all log lines reporting IP assignments with lease durations over a day \\
\texttt{dhcp\_query\_6} & S & W & List all log lines reporting DHCP destination IPs running on non standard ports \\
\texttt{dhcp\_query\_7} & S & W & List all log lines reporting hosts using client version 3.0.1 \\
\texttt{dhcp\_query\_8} & S & W & List all log lines reporting DHCPDISCOVER messages \\
\texttt{dhcp\_query\_9} & S & W & List all log lines reporting XMT Renew messages \\
\texttt{dhcp\_query\_10} & S & W & Return logs reporting bad IP checksums \\
\texttt{dhcp\_select\_1} & S & S & Return DHCP server IP addresses that are not broadcast addresses \\
\texttt{dhcp\_select\_2} & S & S & Find the MAC addresses reported for host laphroaig \\
\texttt{dhcp\_select\_3} & S & S & Return all IPs that were assigned with lease durations over a day \\
\texttt{dhcp\_select\_4} & S & S & Find the unique DHCP destination IPs used with non-standard ports \\
\texttt{dhcp\_select\_5} & S & S & Find the hosts using DHCP client version 3.0.1 \\
\texttt{dhcp\_select\_6} & S & S & List all hosts which sent a DHCPDISCOVER messages \\
\texttt{dhcp\_select\_7} & S & S & Find the unique hosts reporting an XMT Renew message \\
\texttt{simple\_1} & S & W & Show all DHCPACK messages \\
\texttt{simple\_2} & S & W & Find logs from interface eth0 \\
\texttt{simple\_3} & S & W & Show messages from host localhost \\
\texttt{complex\_1} & S & W & Find DHCPREQUEST messages on interface eth0 from host localhost with port 67 \\
\texttt{complex\_2} & S & W & Show DHCPOFFER or DHCPACK messages from server 192.168.1.1 \\
\texttt{complex\_3} & S & W & Find all DHCP messages with transaction ID starting with 0x7 \\
\texttt{vague\_1} & S & W & Show me DHCP errors \\
\texttt{vague\_2} & S & W & Find network problems \\
\texttt{numerical\_1} & S & W & Find entries with process ID greater than 5000 \\
\texttt{numerical\_2} & S & W & Show logs with retry interval less than 5 \\
\texttt{time\_1} & S & W & Show logs from June 18 \\
\texttt{time\_2} & S & W & Find messages from September \\
\texttt{select\_1} & S & S & List all hosts that sent DHCPDISCOVER messages \\
\texttt{select\_2} & S & S & Find the original log timestamps from lines reporting a network is down error, using the timestamp text at the beginning of each matching log line. \\
\texttt{select\_3} & S & S & For each DHCP log line where both are available, return the interface and message type \\
\midrule
\texttt{multiline\_1} & C & S & Find transactions where time from DHCPREQUEST to DHCPACK exceeds 5 seconds. Return a list of pairs of transaction ids and duration. \\
\texttt{multiline\_2} & C & S & Find hosts issuing {$>$}100 DHCPREQUESTs within any 30-minute rolling window. Report the times of the first and last REQUEST of the first such window with the largest number of requests for each host. \\
\texttt{multiline\_3} & C & S & Identify transaction\_ids where a second DHCPREQUEST occurs before the DHCPACK for the first. \\
\texttt{multiline\_4} & C & S & For each client that sent at least one DHCPDISCOVER with an explicit transaction ID, count how many transaction IDs begin with DHCPDISCOVER, how many of those later contain DHCPACK, and the resulting completion rate. \\
\texttt{multiline\_5} & C & S & Considering only client/process streams that contain DHCPDISCOVER, DHCPOFFER, DHCPREQUEST, and DHCPACK, return for each client the average elapsed seconds between the DISCOVER and ACK. \\
\texttt{multiline\_6} & C & S & For each DHCP server IP address, count the total number of OFFER messages sent. Then count how many of those OFFER messages resulted in a corresponding ACK within 60 seconds for the same transaction. Compute the conversion rate. \\
\texttt{multiline\_7} & C & S & For each client, count transaction IDs that contain a DISCOVER and do not contain an ACK within 120 seconds. Only use transactions with an explicit transaction ID, return only clients with a nonzero count, and sort the results in descending order. \\
\texttt{multiline\_8} & C & S & For each client and network interface, count the total number of DISCOVER and REQUEST messages observed. Output one row per client and interface. \\
\texttt{multiline\_9} & C & S & For each client, calculate the average number of DHCP messages logged per transaction ID. Only count DHCP messages that specify a transaction ID.  \\
\texttt{multiline\_10} & C & S & For each hour 0 through 23, count unique transaction IDs whose DISCOVER occurred in that hour. Then count how many of those same transaction IDs have an ACK anywhere in the log. Use only explicit transaction IDs and return all 24 hours with the completion rate for each hour. \\
\texttt{multiline\_11} & C & S & Return the 1-indexed line number and hostname for each DHCPOFFER line where no DHCPDISCOVER from the same hostname/process stream occurred in the previous 30 seconds. If a DHCP message lacks a transaction ID, infer it from the latest transaction ID in the same process stream. \\
\texttt{multiline\_12} & C & S & Identify ACK messages where no matching REQUEST from the same transaction occurred within the previous 30 seconds. \\
\texttt{multiline\_13} & C & S & Identify invalid DHCP cycle transitions by hostname and transaction ID. If a DHCP message lacks a transaction ID, infer it from the latest transaction ID in the same process stream. Valid repeated cycles are DHCPDISCOVER -{$>$} DHCPOFFER -{$>$} DHCPREQUEST -{$>$} DHCPACK and DHCPREQUEST -{$>$} DHCPACK. Treat incomplete trailing sequences as allowed; only explicit bad transitions are violations. \\
\texttt{multiline\_14} & C & S & Identify transactions where two or more REQUEST messages occur before the first ACK. \\
\texttt{multiline\_15} & C & S & For completed transactions, calculate the time difference in seconds between the REQUEST and the ACK. List transactions where this latency exceeds 5 seconds. \\
\texttt{multiline\_16} & C & S & Identify transactions (same client and transaction ID) where two or more DISCOVER messages occur within 60 seconds and no OFFER is present. \\
\texttt{multiline\_17} & C & S & Identify DHCP exchanges where the client received at least two OFFERs from different servers before getting an ACK. Return the client hostname, process id, transaction id, timestamp of the event, list of server ips, and ACKed server for each instance. \\
\texttt{multiline\_18} & C & S & Are there any cases where two different clients were assigned the same IP address at exactly the same timestamp? \\
\texttt{multiline\_19} & C & S & Are there any transactions (same client and transaction ID) where DISCOVER, OFFER, REQUEST, and ACK all occurred at exactly the same timestamp? \\
\texttt{multiline\_20} & C & S & Associate each assigned IP with the MAC address reported for that client/process stream, and return the timestamp, client MAC, and assigned IP. \\
\texttt{multiline\_21} & C & S & Find DHCP renewals where a DHCPREQUEST occurs within 25\% of the stated renewal time after the lease was bound in the same host/process stream. Return the hostname, assigned IP, and elapsed seconds from the bound event to the REQUEST. \\
\bottomrule
\end{tabular}
\end{table*}

\section{Per-Query Results}
\label{app:per-query}
Full per-query $F_1$, precision, recall, TP/FP/FN counts, and generated code for every (dataset, strategy) pair are released with the project artifact.

\section{Example Prompt and Generated Code}
\label{app:prompts}

We reproduce a complete example for sshd complex \texttt{multiline\_5}: ``For each host, count SSH session opens and closes, and compute unclosed sessions.'' The Matryoshka strategy is used with 100 sampled lines. The full prompt (${\sim}$20K tokens) contains: (1) the user query, (2) language constraints (Python 3, stdlib only, no filesystem writes), (3) 102 Matryoshka templates with one example line each, (4) output format and data type specification, (5) 100 sampled lines between \texttt{<SAMPLE>} tags, and (6) a JSON response format instruction. The model returns the following script ($F_1{=}1.000$, TP=123, FP=0, FN=0):

\begin{small}
\begin{verbatim}
import sys, re, gzip
from collections import defaultdict
def parse_logs(filepath):
    stats = defaultdict(lambda: [0, 0])
    pfx = re.compile(
      r'^.{5,40}?\s+(\S+)\s+sshd[^\s:]*:\s')
    opener = (gzip.open
      if filepath.endswith('.gz') else open)
    with opener(filepath, 'rt',
        encoding='utf-8', errors='replace') as f:
      for line in f:
        if 'session opened for user ' in line:
          m = pfx.search(line)
          if m: stats[m.group(1)][0] += 1
        elif 'session closed for user ' in line:
          m = pfx.search(line)
          if m: stats[m.group(1)][1] += 1
    for h,(o,c) in stats.items():
      print(f"{h} {o} {c} {o-c}")
if __name__=='__main__':
    if len(sys.argv)>1: parse_logs(sys.argv[1])
\end{verbatim}
\end{small}

\noindent Despite 102 templates in the prompt, the generated code uses only two substring markers and a syslog-header regex. Full prompts and generated code for all queries are released with the artifact.

\section{Template Strategy Examples}
\label{app:template-examples}

The three auto-generated template strategies produce qualitatively different representations of the same log data. We illustrate using the sshd ``session closed'' event as a running example.

\paragraph{Matryoshka templates} use named semantic placeholders derived from the Matryoshka parser-generation pipeline~\cite{piet2025matryoshka}:
\begin{small}
\begin{verbatim}
TEMPLATE:
  <LOG_TIMESTAMP> <LOG_HOST>
  sshd(<PAM_MODULE>)[<PROCESS_ID>]:
  session closed for user <USER_NAME>
EXAMPLE:
  Sep  4 07:38:02 combo
  sshd(pam_unix)[20024]:
  session closed for user root
\end{verbatim}
\end{small}

\noindent Field names such as \texttt{<USER\_NAME>} and \texttt{<PROCESS\_ID>} encode the semantic role of each variable, providing the model with both the syntactic pattern and its meaning.

\paragraph{Drain3 templates} use generic \texttt{<*>} wildcards, extracted by a streaming fixed-depth prefix tree~\cite{he2017drain}:
\begin{small}
\begin{verbatim}
TEMPLATE:
  Jun <*> <*> <*> <*>
  session closed for user <*>
EXAMPLE:
  Jun  6 15:50:50 orien
  sshd(pam_unix)[25295]:
  session closed for user frank
\end{verbatim}
\end{small}

\noindent Drain3 captures the fixed structure but does not name the variables. The model must infer from context that the first \texttt{<*>} is a day, the second a timestamp, and so on. Drain3 also tends to produce narrower templates (e.g., one per month prefix), resulting in more templates overall.

\paragraph{Frequency templates} use position-based wildcard detection with structural consolidation:
\begin{small}
\begin{verbatim}
TEMPLATE:
  <*> <*> sshd(<*>)[<*>]:
  session closed for user <*>
EXAMPLE:
  Sep  4 07:38:02 combo
  sshd(pam_unix)[20024]:
  session closed for user root
\end{verbatim}
\end{small}

\noindent The frequency templater consolidates templates that share the same structural tokens (here, \texttt{sshd}, \texttt{session closed for user}), absorbing timestamp and hostname variations into a single entry. This produces fewer, broader templates than Drain3, typically at a fraction of the token cost.

\section{Sensitivity to Model Size}
\label{sec:model-compare}

To provide preliminary evidence of generalization across model sizes, we replicate the template-strategy comparison on three datasets (Audit~S, Puppet~C, SSH~C) using Gemini~3 Flash Preview~\cite{gemini3flash2026}, a smaller and less expensive model from the same family. Table~\ref{tab:model-compare} reports the results. The core finding replicates: no single strategy dominates, and auto-generated templates remain competitive with Matryoshka. Flash is comparable to Pro when lightweight templates are provided (Drain3 average 0.871 vs.\ 0.863, Frequency 0.885 vs.\ 0.923), trails on Matryoshka (0.832 vs.\ 0.934), and degrades sharply without context (``none'' drops from 0.655 to 0.586 on average), consistent with the smaller model possessing less pre-trained knowledge of security log formats. Further testing across model vendors (e.g., GPT-5.5, Claude Opus 4.7) remains future work.

\begin{table}[h]
\caption{Pro vs.\ Flash: macro $F_1$ across three datasets and five strategies. \textbf{Bold} = best per row.}
\label{tab:model-compare}
\centering\small
\begin{tabular}{llrrrrr}
\toprule
Dataset & Model & Matry. & Drain3 & Freq. & Rand. & None \\
\midrule
\multirow{2}{*}{Audit (S)}
  & Pro   & 0.984 & 0.978 & 0.978 & \textbf{0.990} & 0.680 \\
  & Flash & \textbf{0.984} & 0.973 & 0.970 & 0.826 & 0.467 \\
\midrule
\multirow{2}{*}{Puppet (C)}
  & Pro   & 0.927 & 0.920 & \textbf{0.945} & 0.745 & 0.684 \\
  & Flash & 0.673 & 0.802 & \textbf{0.835} & 0.638 & 0.520 \\
\midrule
\multirow{2}{*}{SSH (C)}
  & Pro   & \textbf{0.892} & 0.847 & 0.846 & 0.863 & 0.601 \\
  & Flash & 0.839 & 0.837 & \textbf{0.849} & 0.799 & 0.770 \\
\bottomrule
\end{tabular}
\end{table}

\section{Human Baseline Scripts}
\label{app:human}
All \numqueries{} human baseline scripts are available under \texttt{human\_baselines/} in the repository, each annotated with the query it addresses.

\section{Variance Analysis Details}
\label{app:variance}
\label{sec:significance}

To distinguish real effects from LLM non-determinism, we ran $N{=}5$ repeated evaluations with a fixed sampling seed (so that only LLM stochasticity varies) on three simple datasets (Audit~S, SSH~S, Puppet~S) across five strategies and three ablations, yielding 120 evaluations and 1{,}438 per-query $F_1$ measurements. We chose simple datasets because their shorter execution times make 120 evaluations tractable; the noise floor on complex datasets may differ and is left to future work.

The per-query standard deviation across 5 runs (median over queries) sits at 0.000 for almost every (dataset, strategy) cell, with the only exception the ``none'' strategy on Audit~S (median $\sigma{=}0.42$). This means that most queries produce identical $F_1$ across runs. The mean $\sigma$ is higher than the median (Table~\ref{tab:noise-floor}) because a small minority of queries exhibit near-binary behavior, alternating between $F_1{=}0$ and $F_1{=}1$ across runs and pulling the mean upward.

\begin{figure}[h]
    \centering
    \includegraphics[width=\columnwidth]{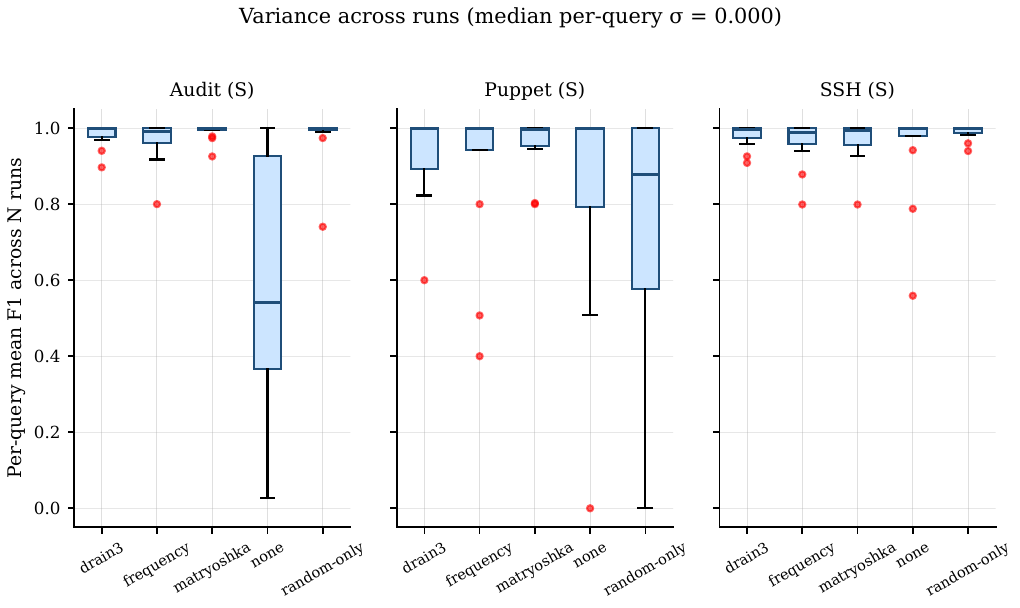}
    \caption{Per-query $F_1$ distributions across $N{=}5$ runs. Template-based strategies are tightly concentrated (median $\sigma < 0.01$); ``none'' shows the widest spread.}
    \label{fig:variance}
\end{figure}

The repeated-run analysis yields three conclusions. First, providing any form of context (templates or samples) produces a large, consistent improvement over no context: an $F_1$ gain of about 0.39 on Audit~S (Matryoshka vs.\ None), with the 95\% confidence interval (Table~\ref{tab:wilcoxon}) excluding zero on every dataset. Second, the differences among template strategies (Matryoshka, Frequency) and between Matryoshka and Random-only fall within run-to-run noise: every pairwise 95\% CI spans zero on every dataset tested. Third, approximately 5\% of queries exhibit per-query $\sigma > 0.4$, oscillating between $F_1{=}0$ and $F_1{=}1$ across runs. These ``binary-failure'' queries represent the boundary of the model's capability: the 4-retry policy partly mitigates this by re-prompting on execution failures, but cannot help when the code runs without error yet returns semantically wrong results.

The implication for the rest of the paper is that template-versus-template comparisons represent directional tendencies within noise rather than definitive rankings. The finding that no strategy dominates is itself robust, because inter-strategy differences are small relative to the gap between any context and no context. The human baseline gaps in \S\ref{sec:baselines} (up to 0.39 $F_1$ on complex datasets) substantially exceed the noise floor.

Table~\ref{tab:noise-floor} reports the per-query noise floor (standard deviation of $F_1$ across runs) for each condition. Table~\ref{tab:wilcoxon} reports all pairwise Wilcoxon signed-rank tests with bootstrap 95\% confidence intervals.


\begin{table*}[!htbp]
\caption{Per-query noise floor: standard deviation of $F_1$ across $N{=}5$ runs. Median $\sigma$ near 0 indicates that most queries produce identical results run-to-run; max $\sigma$ near 0.45--0.55 indicates binary-failure queries that alternate between $F_1{=}0$ and $F_1{=}1$.}
\label{tab:noise-floor}
\centering\small
\begin{tabular}{llrrr}
\toprule
Dataset & Condition & Mean $\sigma$ & Median $\sigma$ & Max $\sigma$ \\
\midrule
\multicolumn{5}{l}{\textit{Template-compare strategies}} \\
Audit (S) & Matryoshka  & 0.001 & 0.000 & 0.012 \\
Audit (S) & Drain3      & 0.049 & 0.000 & 0.447 \\
Audit (S) & Frequency   & 0.051 & 0.003 & 0.447 \\
Audit (S) & Random      & 0.035 & 0.000 & 0.414 \\
Audit (S) & None        & 0.297 & 0.418 & 0.547 \\
Puppet (S) & Matryoshka & 0.078 & 0.000 & 0.447 \\
Puppet (S) & Drain3     & 0.097 & 0.000 & 0.548 \\
Puppet (S) & Frequency  & 0.083 & 0.000 & 0.548 \\
Puppet (S) & Random     & 0.137 & 0.000 & 0.548 \\
Puppet (S) & None       & 0.037 & 0.000 & 0.447 \\
SSH (S) & Matryoshka    & 0.044 & 0.000 & 0.447 \\
SSH (S) & Drain3        & 0.016 & 0.000 & 0.184 \\
SSH (S) & Frequency     & 0.047 & 0.003 & 0.447 \\
SSH (S) & Random        & 0.002 & 0.000 & 0.006 \\
SSH (S) & None          & 0.067 & 0.000 & 0.435 \\
\midrule
\multicolumn{5}{l}{\textit{Prompt-ablation conditions}} \\
Audit (S) & Full         & 0.041 & 0.000 & 0.447 \\
Audit (S) & $-$Sample    & 0.046 & 0.001 & 0.447 \\
Audit (S) & $-$Templates & 0.038 & 0.000 & 0.447 \\
Puppet (S) & Full         & 0.041 & 0.000 & 0.270 \\
Puppet (S) & $-$Sample    & 0.031 & 0.000 & 0.270 \\
Puppet (S) & $-$Templates & 0.083 & 0.000 & 0.447 \\
SSH (S) & Full            & 0.031 & 0.000 & 0.267 \\
SSH (S) & $-$Sample       & 0.053 & 0.000 & 0.546 \\
SSH (S) & $-$Templates    & 0.005 & 0.000 & 0.025 \\
\bottomrule
\end{tabular}
\end{table*}

\begin{table*}[!htbp]
\caption{Pairwise Wilcoxon signed-rank tests on per-query mean $F_1$ ($N{=}5$ runs). $\Delta F_1$ is the mean difference (A $-$ B). 95\% CI from bootstrap. $^{**}p < 0.01$.}
\label{tab:wilcoxon}
\centering\small
\begin{tabular}{llrrl}
\toprule
Dataset & Comparison (A vs.\ B) & $\Delta F_1$ & 95\% CI & $p$-value \\
\midrule
\multicolumn{5}{l}{\textit{Template-compare strategies}} \\
Audit (S) & Frequency vs.\ Matryoshka  & $-$0.024 & [$-$0.060, +0.003] & 0.25 \\
Audit (S) & Frequency vs.\ None        & +0.370 & [+0.189, +0.555] & 0.002$^{**}$ \\
Audit (S) & Frequency vs.\ Random      & $-$0.010 & [$-$0.058, +0.042] & 0.38 \\
Audit (S) & Matryoshka vs.\ None       & +0.394 & [+0.219, +0.575] & 0.002$^{**}$ \\
Audit (S) & Matryoshka vs.\ Random     & +0.015 & [$-$0.003, +0.046] & 1.00 \\
Audit (S) & None vs.\ Random           & $-$0.379 & [$-$0.554, $-$0.210] & 0.004$^{**}$ \\
\midrule
Puppet (S) & Frequency vs.\ Matryoshka  & $-$0.066 & [$-$0.187, +0.035] & 0.77 \\
Puppet (S) & Frequency vs.\ None        & +0.052 & [$-$0.133, +0.269] & 0.62 \\
Puppet (S) & Frequency vs.\ Random      & +0.139 & [$-$0.028, +0.341] & 0.20 \\
Puppet (S) & Matryoshka vs.\ None       & +0.118 & [$-$0.014, +0.311] & 0.45 \\
Puppet (S) & Matryoshka vs.\ Random     & +0.205 & [+0.044, +0.397] & 0.08 \\
Puppet (S) & None vs.\ Random           & +0.087 & [+0.007, +0.183] & 0.12 \\
\midrule
SSH (S) & Frequency vs.\ Matryoshka     & $-$0.007 & [$-$0.056, +0.045] & 0.81 \\
SSH (S) & Frequency vs.\ None           & +0.021 & [$-$0.053, +0.110] & 0.81 \\
SSH (S) & Frequency vs.\ Random         & $-$0.029 & [$-$0.067, $-$0.001] & 0.16 \\
SSH (S) & Matryoshka vs.\ None          & +0.027 & [$-$0.041, +0.110] & 0.94 \\
SSH (S) & Matryoshka vs.\ Random        & $-$0.022 & [$-$0.057, $-$0.002] & 0.05 \\
SSH (S) & None vs.\ Random              & $-$0.049 & [$-$0.132, +0.001] & 0.69 \\
\midrule
\multicolumn{5}{l}{\textit{Prompt-ablation conditions}} \\
Audit (S) & Full vs.\ $-$Sample         & +0.052 & [$-$0.035, +0.168] & 0.55 \\
Audit (S) & Full vs.\ $-$Templates      & +0.001 & [$-$0.047, +0.050] & 1.00 \\
Puppet (S) & Full vs.\ $-$Sample        & $-$0.007 & [$-$0.028, +0.009] & 0.75 \\
Puppet (S) & Full vs.\ $-$Templates     & +0.141 & [+0.018, +0.324] & 0.08 \\
SSH (S) & Full vs.\ $-$Sample           & +0.020 & [$-$0.029, +0.096] & 1.00 \\
SSH (S) & Full vs.\ $-$Templates        & $-$0.016 & [$-$0.041, $-$0.000] & 0.11 \\
\bottomrule
\end{tabular}
\end{table*}

\end{document}